\documentclass[twocolumn]{svjour3}         

\smartqed  

\usepackage{graphicx} 
\usepackage{amssymb}
\usepackage{amsmath}
\usepackage{nicefrac}
\usepackage{color}
\usepackage{sansmath}
\usepackage{physics}
\usepackage{boldline}
\usepackage{multirow}
\usepackage{array}
\usepackage[numbers]{natbib}
\usepackage{hyperref}

\newcolumntype{P}[1]{>{\centering\arraybackslash}p{#1}}

\begin{document}

\title{On the shear-thinning of alkanes}

\author{Hongyu Gao and Martin H. M\"user}

\titlerunning{On the shear-thinning of alkanes}        
\institute{
H. Gao \at
  {
    Dept. Mat. Sci. \& Eng.,
    Saarland University, Campus C6 3,
    66123 Saarbr\"ucken, Germany 
  }
\and M. H. M\"user \at
  {
    Dept. Mat. Sci. \& Eng.,
    Saarland University, Campus C6 3,
    66123 Saarbr\"ucken, Germany 
    \email{martin.mueser@mx.uni-saarland.de}
  }
}

\date{Received: date / Accepted: date}

\maketitle

\begin{abstract}
The approximate power-law dependence of the apparent viscosity of liquids on shear rate is often argued to arise from a distribution of energy barriers.
However, recent work on the Prandtl model, which consists of a point mass being dragged by a damped, harmonic spring past a sinusoidal potential, revealed a similar dependence of the friction on velocity as that of many liquids. 
Here, we demonstrate that this correlation is not only qualitative but can also be made quantitative over a broad temperature range using merely three dimensionless parameters, at least for alkanes, in particular hexadecane, at elevated pressure $p$.
%
%
These and other observations made on our all-atom alkane simulations at elevated pressure point to the existence of an elementary instability causing  shear thinning. 
In addition, the equilibrium viscosity shows power law dependence on $p$ near the cavitation pressure but an exponential dependence at large $p$, while the additional parameter(s) in the Carreau-Yasuda equation compared to other rheological models turn out justifiable. 
%
%
%
\end{abstract}

\section{Introduction}

Research on how liquids oppose shear as a function of shear stress or shear rate, temperature, and also pressure supposedly started almost exactly a century ago~\cite{Barnes1997JNNFM}, when Schalek and Szegvari~\cite{Schalek1923KZ} discovered that the apparent viscosity $\eta$ of iron-oxide colloidal suspension, defined as the ratio of shear stress $\tau$ and shear rate $\dot{\gamma}$, became continuously smaller with increasing shear rate.
%
Ostwald~\cite{Ostwald1925KZ} soon found that this shear thinning, termed ``Strukturviskosität'' at the time (structure or intrinsic viscosity in English), can be approximated rather well with an $\eta ~\propto \dot{\gamma}^{n-1}$ power law, where $n$ is called the shear-thinning exponent.
It is now well established that any liquid, even the prototypical Newtonian fluid water when supercooled~\cite{deAlmeidaRibeiro2020PRR}, has a shear-rate dependent viscosity if the shear rate is only sufficiently high so that labeling fluids as (approximately) Newtonian or not is not a matter of chemistry but mostly of shear rate.  
Liquids obeying Ostwald's empirical law with $n>1$, $n=1$, and $n<1$ can be labeled shear-thickening, Newtonian, and shear-thinning, respectively. 
Under certain situations, $n<0$ might be observable, in which case the friction force decreases with velocity, e.g., in computer simulations of poly-alpha-olefins and 2,4-dicyclohexyl-2-methylpentane at shear rated exceeding 1~GHz~\cite{Ewen2019PCCP}. 
However, macroscopic flow could not be Couette like in such situations, because unavoidable fluctuations of the laminar flow profile would instantly grow, thereby resulting in a dynamical instability producing narrow zones of large slip.
%
%

The equilibrium viscosity $\eta_0(T,p)$ generally increases with pressure $p$ and decreases with temperature $T$~\cite{Bair2003TT,Roland2006JCP}.
One central reason for this behavior certainly is that elevating pressure makes it more difficult for molecules to move past each other, mainly because of increased steric repulsion, while higher temperature assists molecules to overcome energy barriers, whereby shear flow is facilitated. 
The functional form of $\eta(\dot{\gamma})$, e.g., the exponent $n$ in (post-) Ostwald shear-thinning laws, evolves rather smoothly with pressure and temperature with potential exceptions occurring in the vicinity of a liquid-liquid phase transformation.
This has enticed researchers to estimate effective viscosities at temperatures, pressures, and shear rates that are difficult to access experimentally by using time-temperature and pressure superposition principles~\cite{Bair2002PRL}, similar to those used for the construction of master curves describing the linear viscoelasticity of rubbers or glass-forming liquids~\cite{Li2000MSE}. 

It remains actively debated what shear-thinning model describe real-laboratory or \textit{in-silico} experimental data the best~\cite{Ewen2021TL}.
For example, Bair \textit{et al.}~\cite{Bair2015TL} objected to the validity of the Eyring model~\cite{Eyring2004JCP} to describe rheological data, which had been advocated by Spikes and Jie~\cite{Spikes2014TL,Spikes2015TL}.
In a possible retaliation, Spikes~\cite{Spikes2017TL}
dismissed work by Voeltzel, Vergne \textit{et al.}~\cite{Voeltzel2016TL,Voeltzel2017TL} on the rheology ionic liquids with variable shear thinning exponent, by favoring the Eyring model over the originally used Carreau model~\cite{Carreau1972TSR}].
(These and related models are introduced in Sect.~\ref{sec:background}). 
Also the question  what (simple) theory provides the best microscopic explanation for shear thinning in fluids remains discussed, e.g., whether Eyring's~\cite{Eyring2004JCP} or Prandtl's~\cite{Prandtl1928ZAMM} approach is more appropriate~\cite{Spikes2015TL-R}, although it seems often unappreciated that Prandtl had already worked out Eyring theory as a limiting case of his model and that Eyring admitted his results to be similar to Prandtl's and related prior work. 

Another debate evolves around the question what time-temperature, or even time-temperature-pressure superposition principle is best suited for lubricants. 
%
%
For example, Bair~\cite{Bair2017PNAS} objected to the cross-over from a non-Arrhenius to an Arrhenius  dependence of the viscosity upon cooling that Jadhao and Robbins had claimed to have identified for squalane~\cite{Jadhao2017PNAS,Jadhao2017PNAS-R}, though it seems that the transition from non-Arrhenius to Arrhenius dependencies of viscosity and diffusivity, which in some cases is called a fragile-to-strong transition, is a generic phenomenon of glass-forming melts~\cite{Lucas2021JNCS}. 

In this work, we revisit the dependence of the viscosity of simple liquids depends on shear-rate, temperature, pressure and molecular weight.
While numerous simulation studies~\cite{Bair2002PRL,Jadhao2017PNAS,Daivis1994JCP,Sivebaek2012PRL,Lemarchand2015JCP} have computed $\eta(\dot{\gamma})$ relations for realistic lubricant models in the past, often with the attempt to correlate molecular rearrangement with shear thinning~\cite{Lemarchand2015JCP}, see Ref.~\cite{Ewen2021TL} for a recent review, we focus here on the question if the similarity of shear-thinning in realistic model liquids and that observed in the Prandtl model~\cite{Prandtl1928ZAMM} are merely qualitative in nature or if they could be quantitative. 
To this end, we generate a reference relation of $\eta(\dot{\gamma})$ for hexadecane at $T = 500$~K and a hydrostatic pressure of $p = 300$~MPa and identify a Prandtl model that reproduces the $\eta(\dot{\gamma})$ relationship satisfactorily for that reference. 
Next, either temperature or pressure or the molecular weight of the alkane is varied and the changes in equilibrium viscosity studied and discussed in the framework of the Prandtl model, which had been introduced as a simple model for dislocation motion and argued to also explain the velocity dependence of solid friction or the shear thinning of fluids, around similar times when shear thinning was reported for the first time~\cite{Prandtl1928ZAMM,Popov2012ZAMM}.

For our investigation, in which we mean to understand trends rather than to produce accurate numbers for specific molecules, 
we see computer simulation as the ideal tool, despite essentially unavoidable flaws in interaction potentials and the difficulty to predict reliable viscosity data for shear rates below 10$^7$~Hz. 
However, the advantage of simulations is that high shear rates can be reached while conditions such as shear rate, pressure, and temperature can be set to a target precision.
Moreover, undesired effects like frictional heating can be suppressed or at least be quantified. 
Last but not least, the pressure dependence of the viscosity can be investigated even under a well-defined isotropic, tensile stress.
%

%
%


%

We continue this paper with a background section on rheological models in Sect.~\ref{sec:background}.
Theory and numerical methods are introduced in Sect.~\ref{sec:theory}.
Results are presented in Sect.~\ref{sec:results}, while conclusions are drawn in Sect.~\ref{sec:conclusions}.

\section{Background on rheological models}
\label{sec:background}

%
%
The rheology of a broad class of different liquids including standard base oils, water, or blood is
 captured quite well by the Carreau-Yasuda (CY) equation, 
\begin{equation}
\label{eq:CarreauYasuda}
    \eta(\dot{\gamma}) = \eta_\infty + (\eta_0-\eta_\infty)\left\{1+\left(\dot{\gamma}/\dot{\gamma}_0\right)^a
    \right\}^{(n-1)/a},
\end{equation}
where $\eta_0$ is the equilibrium viscosity, $\dot{\gamma}_0$ is a reference shear rate, near which shear thinning starts or has started to set in, while $\eta_\infty$ is a parameter discussed in detail in the next paragraph. 
Two other popular models arise as limiting cases of Carreau-Yasuda, namely the Carreau model~\cite{Carreau1972TSR} for $a=2$ and the Cross model~\cite{Crook1963MPS} for $a=1-n$.
%

An arguably tricky term in Eq.~(\ref{eq:CarreauYasuda}) is the large-shear-rate limit of the viscosity, $\eta_\infty$, 
%
even if it is often useful to introduce it
in practice so that fitting functions extend their validity to large shear rates.
However, at the point at which $\eta(\dot{\gamma})$ appears to level off, the way in which stress and temperature are controlled is likely to matter.
%
%
For example, we find that fitting viscosity data when reported at fixed pressure does not require $\eta_\infty$ to be introduced, while it does at constant density. 
These observations are in line with classical results by Daivis and Evans~\cite{Daivis1994JCP}, who found decane to show a finite value of $\eta_\infty$ or even to shear thicken at constant-density simulations but not at constant pressure.
For these reasons, $\eta_\infty$ will be ignored in the following, all the more extreme shear-thinning at very large $\dot{\gamma}$ is readily built in and thus rationalized in the Prandtl model, see Sect.~\ref{sec:prandtl}.

Coming back to the discussion of rheological models:
Carreau~\cite{Carreau1972TSR} had originally attempted to simultaneously address the linear-response viscoelasticity and the out-of-equilibrium shear thinning by assuming the rates of creation and annihilation of molecular junctions to be functions of what now would be called the equivalent von Mises strain-rate. 
Carreau suggested two \textit{ad-hoc} functional dependencies for those functions, one of which he dismissed as unrealistic, which, ironically is the one leading to the equation now carrying his name, i.e., Eq.~\eqref{eq:CarreauYasuda} for $a = 2$. 

A beneficial aspect of the Carreau model is that the leading-order corrections to the linear-response viscosity are, as they should be, of order $\dot{\gamma}^2$~\cite{Muser2020L}. 
However, it seems that expanding the Carreau equation into (even) powers of the shear rate does not accurately reflect the initial but only the later deviations of the viscosity from the equilibrium viscosity.
We base this conclusion on the observation that almost any high resolution data that we are aware of, be it experimental or simulations, lies below the fits at the point where $\eta$ has decreased to say 80\% its equilibrium value.
%

%

Another shear-rate dependence of the viscosity, which can be motivated from Eyring theory, is
\begin{equation}
\label{eq:Eyring}
    \frac{\eta}{\eta_0} = \frac{\dot{\gamma}_0}{\dot{\gamma}}\mathrm{asinh}\left(\frac{\dot{\gamma}}{\dot{\gamma}_0}\right),
\end{equation}
where the ideal value for $\dot{\gamma}_0$ may differ from that of another rheological model. 
This relation makes the shear stress increase logarithmically with $\dot{\gamma}$ at large  $\dot{\gamma}$ while converging to the equilibrium viscosity $\eta_0$ at $\dot{\gamma} \to 0$, as is typical for boundary lubrication.
This is arguably the main reason for the use of this and related relations 
in state-rate models of frictional contacts~\cite{Ruina1983JGR}. 
%
%
However,  what appears to be often overlooked is that Eq.~\eqref{eq:Eyring} can be seen as a limiting case of Eq.~\eqref{eq:CarreauYasuda}, because an increase of shear stress or friction with a power-law $\dot{\gamma}^\varepsilon$ with $0 < \varepsilon \ll 1$ behaves essentially logarithmically.
%
%
%
Thus, as long as $0 < n \ll 1$, discrepancies between Eyring and the other rheological models can only be minor, in which case the only advantage of Eyring would be to depend on one less adjustable parameter. 
%


%

\begin{figure}[hbtp]
\centering
\includegraphics[width=0.475\textwidth]{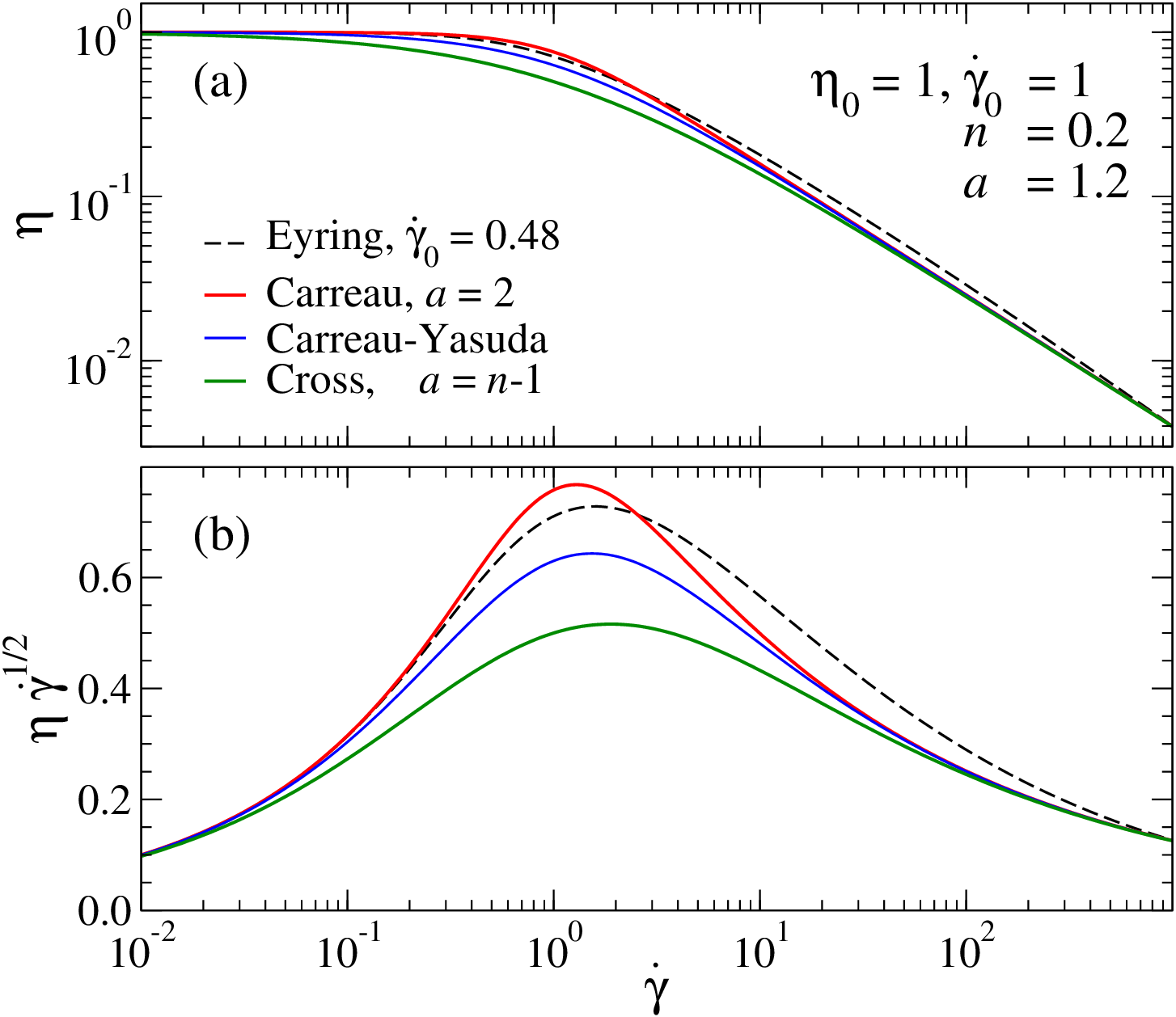}
\caption{Shear-rate dependence of the viscosity $\eta$ as produced by various model, (a) in a regular representation and (b) in an alternative fasion.
All model functions assume the same linear-response viscosity $\eta_0$ and a similar $\eta(\dot{\gamma}=10^3)$. Carreau-Yasuda (blue line), Carreau (red), and Cross (green) employ the same value for $n$ and $\dot{\gamma}_0$.} 
\label{fig:CYetAl}
\end{figure}

When evaluating the functional form of a given model, mostly the viscosity and sometimes the shear stress or friction are often represented in a double logarithmic plot as a function of shear rate or velocity.
This way differences between different models near the cross-over from Stokes friction to ``Ostwald scaling'' have a low optical resolution.
It can be improved, in particular for shear-thinning exponents near ``Eyring scaling'', i.e., $0 < n \ll 1$, when multiplying the viscosity with the square-root of the shear rate~\cite{Muser2011PRB,Muser2020L}.
An example of such a  what-we-take-the-liberty-to-call-for-reasons-of-simplicity-and-not-vanity Müser plot is shown in  Fig.~\ref{fig:CYetAl}.
It contrasts the regular $\eta(\dot{\gamma})$ representation for the exponent $n=0.2$ with a Müser plot. 
In this example, $n$ was chosen roughly half way between the values identified in this study and $n=1-\varepsilon$, $0<\varepsilon \ll1$, at which point the algebraic dependence of shear stress on shear rate is logarithmic like. 

To quantify the goodness of fits to molecular dynamics (MD) results, we defined a $\chi^2$-error as a logarithmic deviation of the viscosity function from the reference data, i.e.,
\begin{equation}
\chi^2 = \frac{1}{N}\sum_{n=1}^N 
\left\{ \ln\frac{\eta_\textrm{model}(\dot{\gamma}_n)}{\eta_\textrm{MD}(\dot{\gamma}_n)}
\right\}^2,
\end{equation}
where $N$ enumerates the data points.
This way, errors do not depend on whether stress or viscosity are targeted.
To not bias the fits to large or low shear rates, we furthermore used $\dot{\gamma}_{n+1}/\dot{\gamma}_n$ constant.  

\section{Theory and methods}
\label{sec:theory}

\subsection{Inertia-free Prandtl model}
\label{sec:prandtl}
 
In the Prandtl model, a mass point is pulled with a spring over a corrugated substrate.
In addition to the conservative forces, Stokes' friction terms arise, which are argued to reflect lattice vibrations in the substrate.
However, this explicit damping can also be assumed to occur in the pulling spring, in which case the mass point's equation of motion in the presence of thermal noise $\Gamma_\textrm{th}(t)$ reads
\begin{equation}
\gamma_0 (\dot{x}-v_0) = qV_0\sin(qx) + k (x-v_0 t) + 
\Gamma_\textrm{th}(t)
\end{equation}
Here $x$ is the position of the ``Prandtl atom'', $\dot{x}$ its velocity, $v_0$ the velocity of the pulling spring, $\gamma_0$ is a damping term of unit kg/s, $V$ the amplitude of the sinusoidal potential through which the atom is dragged, $2\pi/q$ is the period of the potential, $k$ the stiffness of the pulling spring, and $\Gamma_\textrm{G}$ is thermal noise with the expectation values
\begin{eqnarray}
\langle \Gamma_\textrm{th}(t) \rangle & = & 0 \\
\langle \Gamma_\textrm{th}(t) \Gamma_\textrm{th}(t') \rangle & = &  2\gamma_0 k_\textrm{B}T \delta(t-t'),
\end{eqnarray}
where $k_\textrm{B}T$ the thermal energy and $\delta(...)$ is the Dirac delta function.
We have the liberty to use a unit system of our choice, i.e., one in which $q$, $V_0$, and $\gamma$ define the units used for inverse length, energy, and weight per time. 
When doing so, it becomes clear that this form of the Prandtl model has two parameters defining the shape of a $F(v)$ relation, since only the reduced stiffness $\tilde{k}\equiv k/(q^2V)$ and the reduced temperature $\tilde{T}\equiv k_\textrm{B}T/V$ remain, in which case the dynamical equation to be solved, introducing $\langle \tilde{\Gamma}_\textrm{th}(t) \tilde{\Gamma}_\textrm{th}(t')\rangle =\delta(t-t')$, simplifies to
\begin{equation}
\dot{x}-v_0 = \sin x + \tilde{k}(x-v_0 t) + \sqrt{2\tilde{T}}\tilde{\Gamma}_\textrm{th}(t),
\end{equation}

\begin{figure}[hbtp]
\centering
\includegraphics[width=0.4\textwidth]{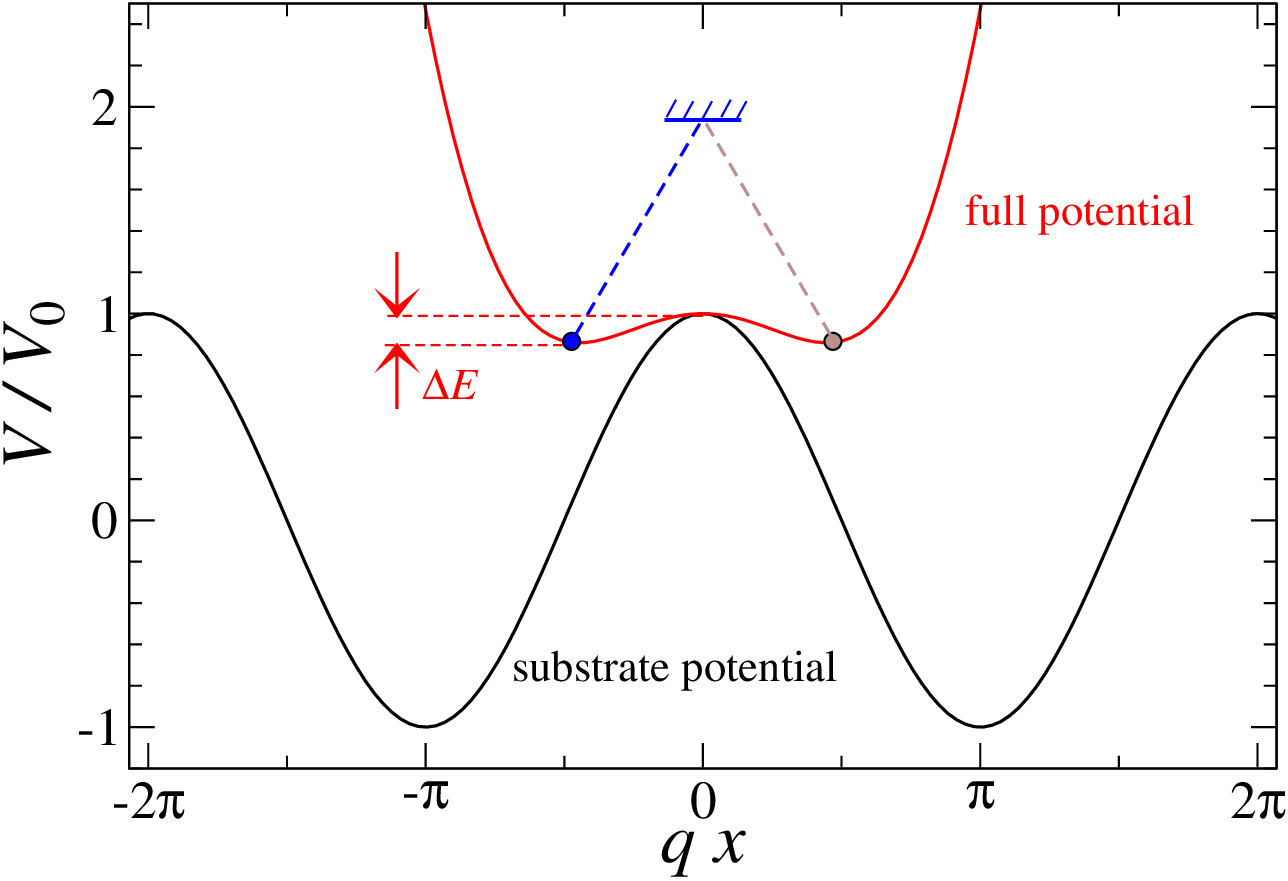}
\caption{Substrate and total potential energy for $\tilde{k} = 0.7$, which is used at the reference state (300~MPa, 500~K, which corresponds to $\tilde{t} = 0.076$). 
{The energy barrier $\Delta E=0.14$} is defined at the moment of time when the position of the driving spring coincides with the maximum of the substrate potential. 
}
\label{fig:potential_landscape}
\end{figure}

In this model, only two dimensionless numbers, $\tilde{k}$ and $\tilde{T}$, determine the \textit{shape}  of a $\eta(\dot{\gamma})$ relation, e.g., the exponents $n$ and $a$ that would be obtained when fitting simulation data to a Carreau-Yasuda relation.
Absolute values, such as the equilibrium viscosity $\eta_0$ or the cross-over shear rate $\dot{\gamma}_0$ are then determined by the model parameters defining the units in the Prandtl model. 
For example, given a value of $\tilde{k}$ and assuming $\tilde{k}$ and $V_0$ to be temperature-independent, the small-temperature dependency of the equilibrium-damping constant of the Prandtl model, which is proportional to the viscosity, reveals Arrhenius type dependence~\cite{Muser2020L} with the energy activation barrier $\Delta E$ whose construction is  shown in Fig.~\ref{fig:potential_landscape}.
Assuming the temperature to be below the liquid's fragile-to-strong transition temperature and $\Delta E$ to be know, the unit for $V_0$ can then be fixed.

When varying temperature of a large range, it seems clear that both $V_0$ and $k$ depend on temperature.
Specifically, at large $T$, the all-atom liquid expands, which will reduce $V_0$ supposedly more than $k$.
To reflect such effects of thermal dilation in the model, we allowed $\tilde{k}$ to obey
\begin{equation}
\tilde{k} = \tilde{k}_\infty \exp(-T_k/T),
\label{eq:springFudge}
\end{equation}
where $T_k$ is an empirical parameter, which would need to get readjusted as a function of pressure and the degree of polymerization. 
The exponential dependence is motivated by the Boltzmann distribution without any further justification.
However, we note that the value used for $T_k$ was only $T = 90.5$~K, is clearly less than the smallest investigated temperature of 300~K.

Before discussing some important asymptotic limits of the Prandtl model, we wish to justify why we damp the spring motion rather than the velocity of an atom or rather a material point relative to the substrate.
First, the kinetic energy of monolayers sliding across a substrate was shown to be related to the (anharmonic) coupling of vibrations in the monolayer~\cite{Smith1996PRB} and we see a Prandtl spring to reflect the coupling of a central atom to its in-plane neighbors of a lamella during laminar flow. 
Second, damping relative to the substrate would yield a high-velocity damping, which would produce a finite $\eta_\infty$.
However, as discussed in Sect.~\ref{sec:background}, we see no evidence for the existence of $\eta_\infty$, when evaluating $\eta(\dot{\gamma})$ at a constant pressure. 

A few properties of the Prandtl model are worth reciting, most of which can be found in~\cite{Fisher1985PRB}.
First, instabilities and thus (athermal) static friction and Coulomb-like friction at low temperatures can only occur when $\tilde{k}<1$.
Second, the athermal low-velocity limit of the friction-velocity relation reads $F_\textrm{k}-F_\textrm{s} \propto v^{2/3}$.
Thus, for $\tilde{k}$ being in the immediate vicinity of unity, a shear-thinning exponent of $n=2/3$ is obtained.
This value thereby constitutes an upper bound for what shear-thinning exponent $n$ can be reproduced with the Prandtl model.
Third, in the limit $0 < \tilde{k} \ll 1$, the material point is effectively moved with a constant force, which is the limit at which the Prandtl model corresponds to situation described by himself and later again by Eyring.
In that limit, the effective damping at small but finite temperature can be described with Eq.~\eqref{eq:Eyring}.
Thus, the Prandtl model can reproduce any shear-thinning exponent between $0 < n \le 2/3$, where a small exponent would imply small $\tilde{k}$ and/or finite temperature, while $n=2/3$ would suggest thermal fluctuations to be minor and/or $\tilde{k}$ to be relatively close to unity. 
In fact, the Prandtl model reproduces the general trend of real lubricants~\cite{Ewen2021TL} that $n$ decreases with decreasing temperature~\cite{Muser2020L}.

The fourth asymptotic limit to be discussed is the high-velocity limit.
In this case, the atom moves at constant velocity of $v_0$ plus small fluctuations, which can be described using perturbation theory,
The first-order perturbation is a time-dependent force $qV\sin(\omega t)$ with $\omega = qv_0$.
For $\omega\to\infty$, the response function $\tilde{x}_1(\omega)~\propto 1/\omega$ so that the dissipated power $P \propto \vert \omega \tilde{x}_1 \vert^2$ is independent of $\omega$, which in turn means that the friction force is proportional to $1/\omega$ or $1/v$, which, when applied to laminar flow would translate to $n=-1$. 
%
Couette flow would no longer be possible, as discussed in the introduction.
Adding inertia to the model would further impede Couette flow at large $\dot{\gamma}$ as $\tilde{x}_1$ would then be proportional to $1/\omega^2$ at large $\omega$, which decreases $n$ to $n=-3$.
Irrespective of the choice of the mass, extreme shear thinning at very high shear rate occurs in the Prandtl model, when the spring head moves so fast that the energy released during an instability can no longer be transferred into heat before the atom undergoes the next instability. 
We would argue that similar statements can be made for any other systems and degrees of freedom showing decreasing friction with increasing velocity.

\subsection{Molecular-dynamics alkane simulations}

Non-equilibrium molecular dynamics (NEMD) simulations are carried out on a bulk liquid consisting of approximately 5{,}000 atoms using the open-source code--LAMMPS~\cite{Plimpton1995JCP}.
Interatomic interactions are modeled with the all-atom L-OPLS-AA force field~\cite{Price2001JCC,Siu2012JCTC}, which describes bond-stretching, bending, and torsional in addition to non-bonded interactions.
The shear viscosity is calculated through Couette-flow simulations in which shear flow is implemented via Lees-Edwards~\cite{Lees1972JPC} periodic boundary conditions and the  SLLOD algorithm \cite{Todd2017}.
%
%
%
The system temperature, $320~\textrm{K} \le T \le 550$~K was maintained with a Nos\'e-Hoover thermostat. 
To this end, the LAMMPS command temp/deform was used that removes the effect of a mean velocity profile from the determination of kinetic energy and thus temperature. 

%
Results are reported for a fixed normal pressure, which was achieved as follows:
In a first set of short simulations at fixed volumes and shear rates, an equation of state for the give shear rate was ascertained and the volume  determined by interpolation (never extrapolation!) at which the target pressure was expected to occur.
This volume was then used in a longer run during which the effective viscosity was calculated. 
The average normal stress in those long runs always turned out within 1\% at finite stress and $\pm 1$~MPa at zero stress, where small relative errors are somewhat difficult to achieve. 
In the case of moderate, mean tensile stresses, the same average value $\sigma_{33}$ value can be reached with two different densities. 

\section{Results}
\label{sec:results}

\subsection{Temperature dependence}

\begin{figure}[hbtp]
\centering
\includegraphics[width=0.4\textwidth]{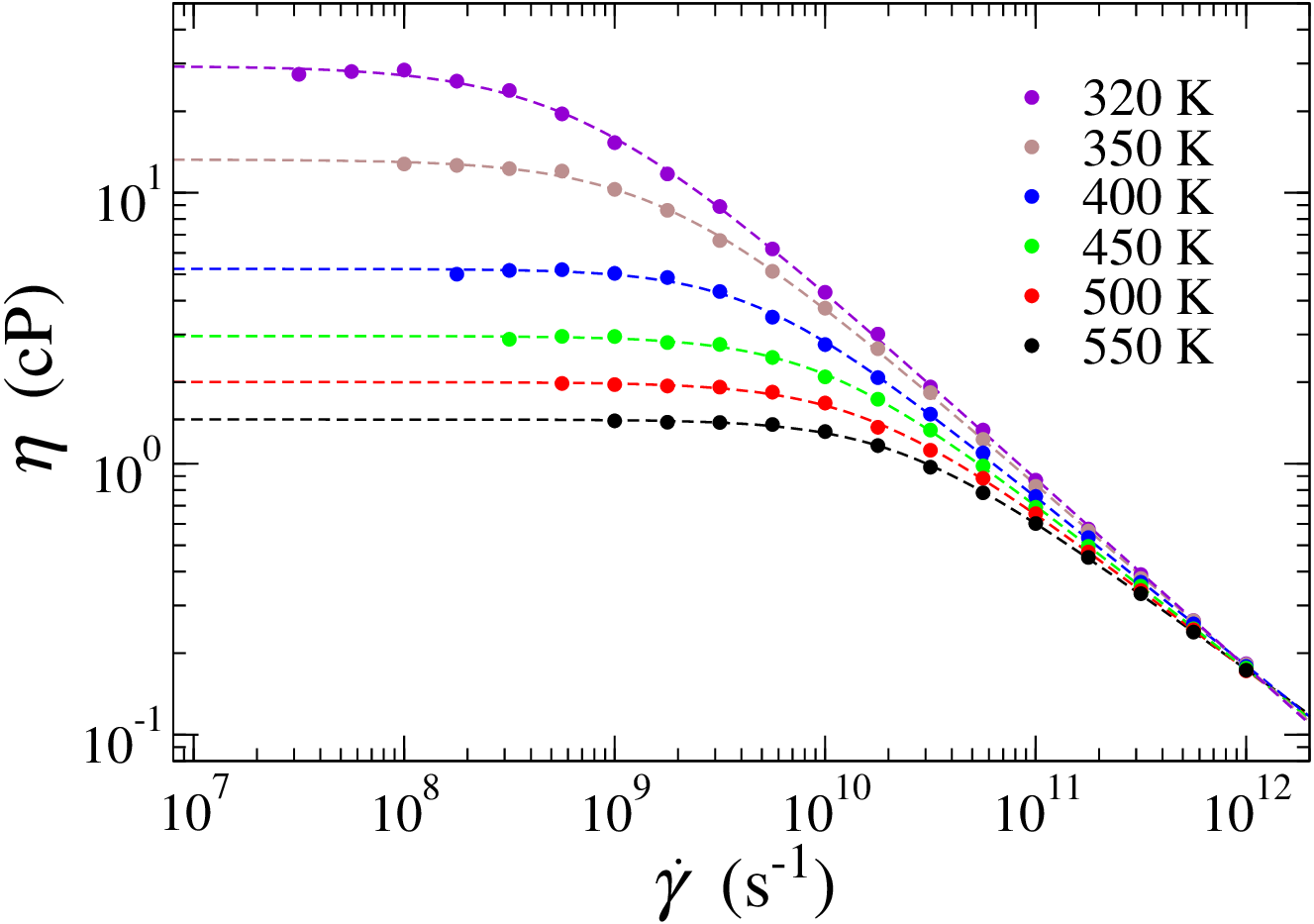}
\caption{Shear viscosity $\eta$ of $n$-hexadecane at various temperatures (320-550 K) and a constant normal pressure of 300 MPa. Symbols show NEMD data and dashed lines fits to the data using the Carreau-Yasuda equation.}
\label{fig:eta_gamma_temp}
\end{figure}

We first investigate the dependence of the viscosity of hexadecane as a function of temperature at a pressure of $p_{33} = 300$~MPa, where $p_{33} \equiv -\sigma_{33}$.
The latter value was chosen such that it is high enough to almost represent boundary-lubrication conditions but small enough so that the equilibrium viscosity could be calculated down to a temperature of $T = 320$~K, which would be at the cold end of engine oil after warming up. 
%
%
Fig.~\ref{fig:eta_gamma_temp} shows the raw data obtained from MD simulations and fits based on the Carreau-Yasuda equation. 

The range of shear rates presented in Fig.~\ref{fig:eta_gamma_temp} and others presented in this work is certainly beyond any practical relevance, which should be roughly $\dot{\gamma} = 1$~GHz, obtained when two surfaces moving at a relative velocity of 2~m/s have a local separation of 2~nm.
However, owing to time-temperature and time-pressure superposition principles, similar rheology (except for prefactors) can occur at ``reasonable'' shear rates when the temperature is lowered and/or the pressure increased.
Nonetheless, using shear-rates clearly above $10^{12}$~Hz would not be meaningful, because these rates exceed local librational frequencies. 

\begin{figure}[hbtp]
\centering
\includegraphics[width=0.4\textwidth]{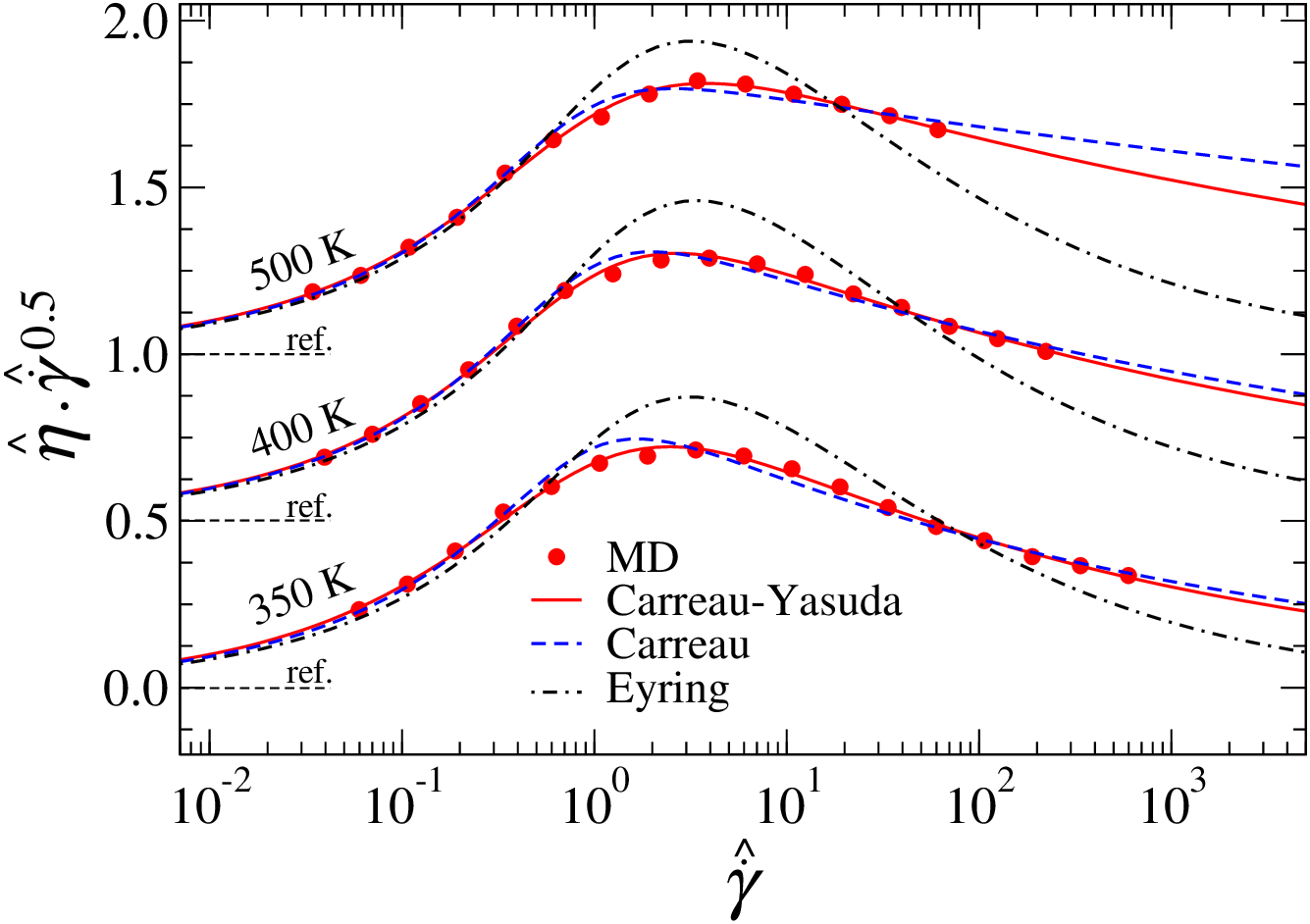}
\caption{\label{fig:mueser_plot_waterfall}
A comparison of phenomenological models fitted to a representative case of $n$-hexadecane at $T$ = 350, 400, and 500 K and $p_{33}$ = 300 MPa with normalized $\hat{\dot\gamma}=\dot\gamma/\dot\gamma_0^\textrm{CY}$ and $\hat{\eta}=\eta/\eta_0^\textrm{CY}$, where the upper index CY indicates that the Carreau-Yasuda values at the given values were taken as units for viscosity and frequency. Results from $T = 400$~K and 500~K are shifted in $y$ direction by 0.5 and 1.0, respectively.
}
\end{figure}

Although it has been a matter of ongoing debates what functional form describes data as that presented in Fig.~\ref{fig:eta_gamma_temp} best, we feel that 
%
a ``Müser plot''
allows differences between fit and data to be much better resolved than traditional representations of $\eta(\dot{\gamma})$. 
%
%
%
For the data in question, our results for viscosity spanning two decades in a regular $\eta(\dot{\gamma})$ representation as in Fig.~\ref{fig:eta_gamma_temp} vary merely by a factor of about three in the cross-over domain, as can be seen in Fig.~\ref{fig:mueser_plot_waterfall}.
This improves the resolution of the graph effectively by a factor of approximately 60, whereby both small deficiencies of rheological model or stochastic errors are relentlessly revealed.

Fig.~\ref{fig:mueser_plot_waterfall} compares three rheological models for the viscosity of hexadecane at three different temperatures.
It can be seen by the naked eye that the Eyring model produces the data in the least satisfactory manner.
Carreau is a substantial improvement to Eyring, however, Carreau overestimates the viscosity at shear rates slightly lower than the maximum.
%
We invite the reader to check for this deviation to occur in other published data and invite him or her to remember our assessment that this short-coming of Carreau shows up in the early shear thinning regime the more clearly the less noisy the real-laboratory or \textit{in-silico} data. 

%
%
%
%
Relative standard-deviations converted in percentages, $\chi^\%\equiv 100 \cdot  \chi$, are listed in
Table~\ref{tab:CY_fit_results} for all rheological models introduced in this paper so far and all MD-based $\eta(\dot{\gamma})$ reference data.
%
The latter includes data obtained by varying the pressure or the molecular weight rather than the temperature from their default values. 

\begin{table*}[thbp]
\centering
\begin{tabular}{P{0.5cm}|P{0.8cm}|c|c|c|P{0.8cm}|P{0.7cm}|P{0.8cm}|P{0.8cm}|P{0.8cm}|P{0.8cm}} \hlineB{3}
$P$ & $T$/K & $p_3$/MPa & $\eta_0$/cP & $\dot{\gamma}_0$/GHz & $n$ & $a$ & $\chi_\textrm{CY}^{\%}$ & $\chi_\textrm{C}^{\%}$ & $\chi_\textrm{Cr}^{\%}$ & $\chi_\textrm{E}^{\%}$\\ \hline
\multirow{6}{*}{16} & 320 & \multirow{6}{*}{300} & 26.76 & 0.71 & 0.310 & 1.28 & 4.12 & 5.45 & 6.46 & 17.77\\
 & 350 & & 13.25 & 1.67 & 0.327 & 1.17 & 1.93 & 4.11 & 4.56 & 15.78\\
 & 400 & & 5.24 & 4.50 & 0.376 & 1.42 & 1.86 & 2.79 & 5.41 & 13.73\\
 & 450 & & 2.96 & 9.65 & 0.391 & 1.34 & 1.25 & 2.67 & 4.53 & 11.53\\
 & 500 & & 2.00 & 16.37 & 0.406 & 1.25 & 1.65 & 3.00 & 3.71 & 10.64\\  
 & 550 & & 1.46 & 22.35 & 0.441 & 1.36 & 0.97 & 2.06 & 3.51 & 9.96\\  \hline
\multirow{6}{*}{16} & \multirow{6}{*}{500} & 100 & 0.66 & 59.25 & 0.300 & 1.16 & 1.61 & 3.29 & 2.41 & 4.66\\
 & & 200 & 1.20 & 26.37 & 0.404 & 1.35 & 1.67 & 2.58 & 3.69 & 8.03\\
 & & 300 & 2.00 & 16.37 & 0.406 & 1.25 & 1.65 & 3.00 & 3.71 & 10.64\\  
 & & 400 & 2.97 & 9.66 & 0.425 & 1.45 & 1.74 & 2.44 & 5.00 & 12.63\\
 & & 500 & 4.36 & 7.61 & 0.398 & 1.26 & 1.84 & 3.16 & 4.38 & 13.47\\
 & & 600 & 5.90 & 5.31 & 0.404 & 1.45 & 1.07 & 2.14 & 5.45 & 14.46\\ \hline
16 & \multirow{3}{*}{500} & \multirow{3}{*}{300} & 2.00 & 16.37 & 0.406 & 1.25 & 1.65 & 3.00 & 3.71 & 10.64\\  
20 &  &  & 2.81 & 8.54 & 0.421 & 1.32 & 1.77 & 2.88 & 4.61 & 13.62\\      
24 &  &  & 3.83 & 5.26 & 0.418 & 1.19 & 1.97 & 3.53 & 4.51 & 16.26\\  \hlineB{3}
\end{tabular}
\caption{Results for Carreau-Yasuda (CY) fits to MD data. $P$ stands for the number of backbone carbons. Relative standard deviations are also included for fits to Carreau (C), Cross (Cr), and Eyring (E). 
}
\label{tab:CY_fit_results}
\end{table*}

Since Carreau (Ca) and Cross (Cr) are exact limits of Carreau-Yasuda (CY), they cannot outperform CY. 
It is similarly clear that Eyring (E) should be the least accurate of the four models, because it arises as an approximate asymptotic limit from the other models. 
However, the interesting question is to what degree increasing the number of fit parameters from 2 (E) to 3 (Ca, Cr) to 4 (CY) improves the goodness of the fits. 
Our personal view is that the dividing line between justified and unjustified additional parameter is a reduction of $\chi$ by a factor that should be clearly smaller than $k/(k+1)$ when $k$ is the number of originally adjustable coefficients. 

Table~\ref{tab:CY_fit_results} reveals quite clear trends.
Errors  associated with Eyring exceed 10\% in all but two out of 12 sets of simulations.
The Cross model reduces the errors compared to Eyring by a factor always exceeding 1.9 and in the clear majority of cases by a factor greater than 2.5.
Thus, the additional adjustable parameter seems justified.
Carreau further reduces the error compared to Cross quite substantially, except at the highest temperature and the lowest pressure reported in Table~\ref{tab:CY_fit_results}.
Carreau-Yasuda is again a clear improvement on Carreau.
Improvements scatter between reduction factors of 1.3 and 2.2, however, the median improvement in the reported data lies at about 1.8.
For this reason, we judge the extra parameter $a$ in Carreau-Yasuda as significant, in particular as the Carreau-Yasuda errors are often only marginally larger than the statistical uncertainty of the produced data. 
Unfortunately, it is beyond our current resources to substantially reduce the thermal/statistical errors, because halving statistical errors implies quadrupling the numerical effort.

While all investigated model functions, even that based on Eyring, matches the MD data in the representation of Fig.~\ref{fig:mueser_plot_waterfall} substantially better than an ``arbitrary fitting function"---such as a skewed Gaussian, which, like Carreau-Yasuda,  has four adjustable parameters---Cross, Carreau, and Carreau-Yasuda are purely heuristic functions with built-in linear response at small shear rates and power-law dependence at large shear rates.
However, unlike Eyring and even more so Prandtl, they lack a clear microscopic justification.  
The Prandtl model is a physics-based model, in which linear response at small velocities arises naturally while quasi-power-law behavior occurs at large shear rates~\cite{Popov2012ZAMM,Muser2020L}. 

The just-mentioned feature of the Prandtl model motivated us to explore to what extent the two dimensionless parameters of the Prandtl model can be tuned to match experimental or realistic, \textit{in-silico} data.
The interesting aspect of such an analysis is that the model parameters can be fixed at one temperature and changes in $\eta(\dot{\gamma})$ be estimated without requiring adjustment of the parameters beyond including a smooth temperature dependence of $\tilde{k}$.

\begin{figure}[hbtp]
\centering
\includegraphics[width=0.4\textwidth]{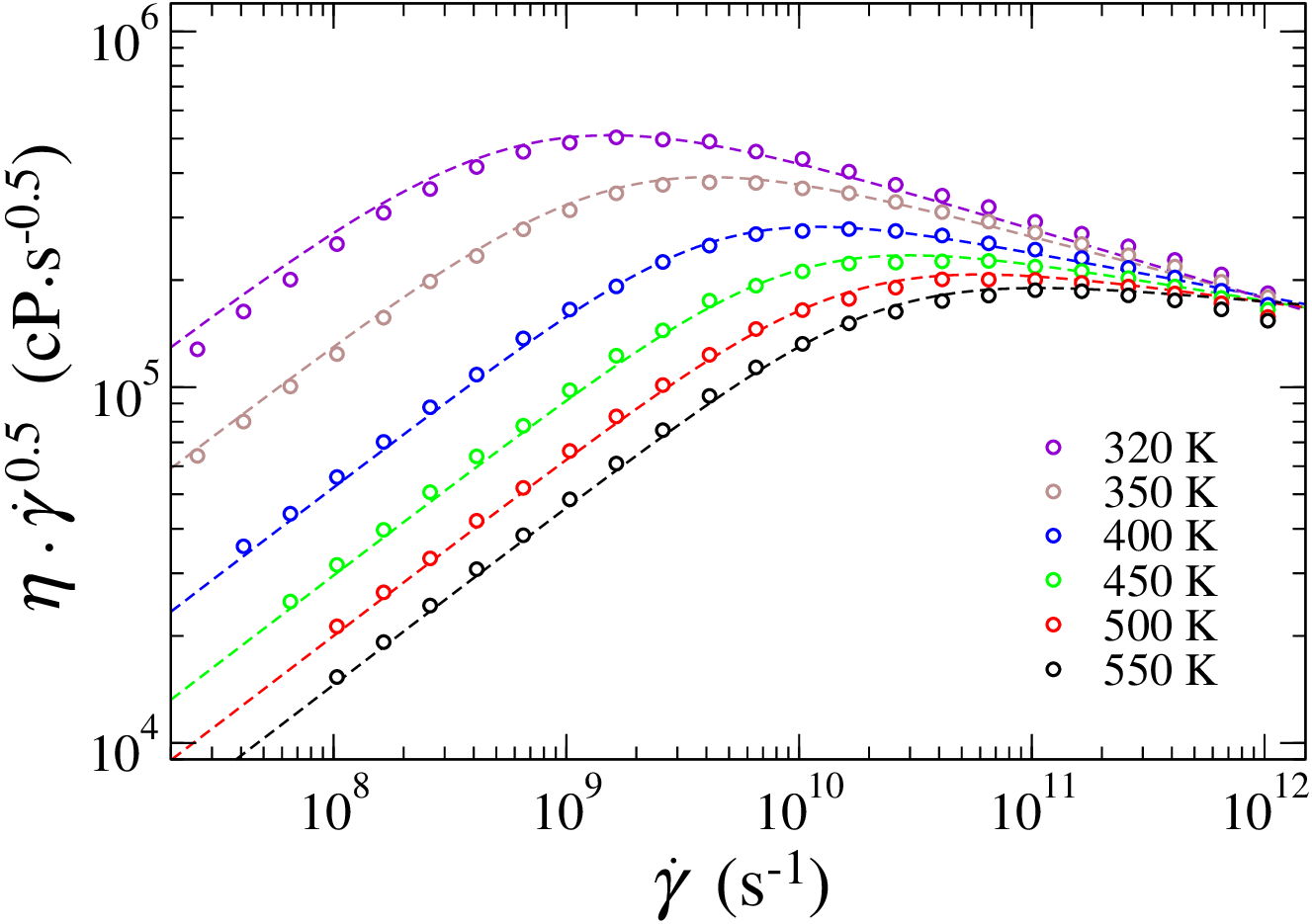}
\caption{\label{fig:Prandtl_MD_mapping}
Comparison of results obtained from the Prandtl model to Carreau-Yasuda fits to the hexadecane MD results at $p_{33} = 300$~MPa (dashed curves). The Prandtl model was chosen with 
$\tilde{k}=0.84\cdot(1 - 90.5~K/T)$ and $\tilde{t} = 1.52\cdot 10^{-4}~T$/K. 
}
\end{figure}

Fig.~\ref{fig:Prandtl_MD_mapping} reveals that three dimensionless parameters and thus a total of five parameters is sufficient to reproduce the full-scale NEMD results at different temperatures. 
Interestingly, it also reproduces the observation that the exponent $n$ decreases thereby approaching Eyring-like behavior with decreasing temperature.
Specifically, the full simulations show $n = 0.391$ at $T = 450$~K and $n = 0.327$ at $T = 350$~K, while the Prandtl model finds $n = 0.366$  and $0.321$ at those two temperatures, respectively.
%
%
Thus, the Prandtl model does not perfectly reproduce the reduction in the shear-thinning exponent  but supposedly much better than expected from a simple model.

\begin{figure}[hbtp]
\centering
\includegraphics[width=0.4\textwidth]{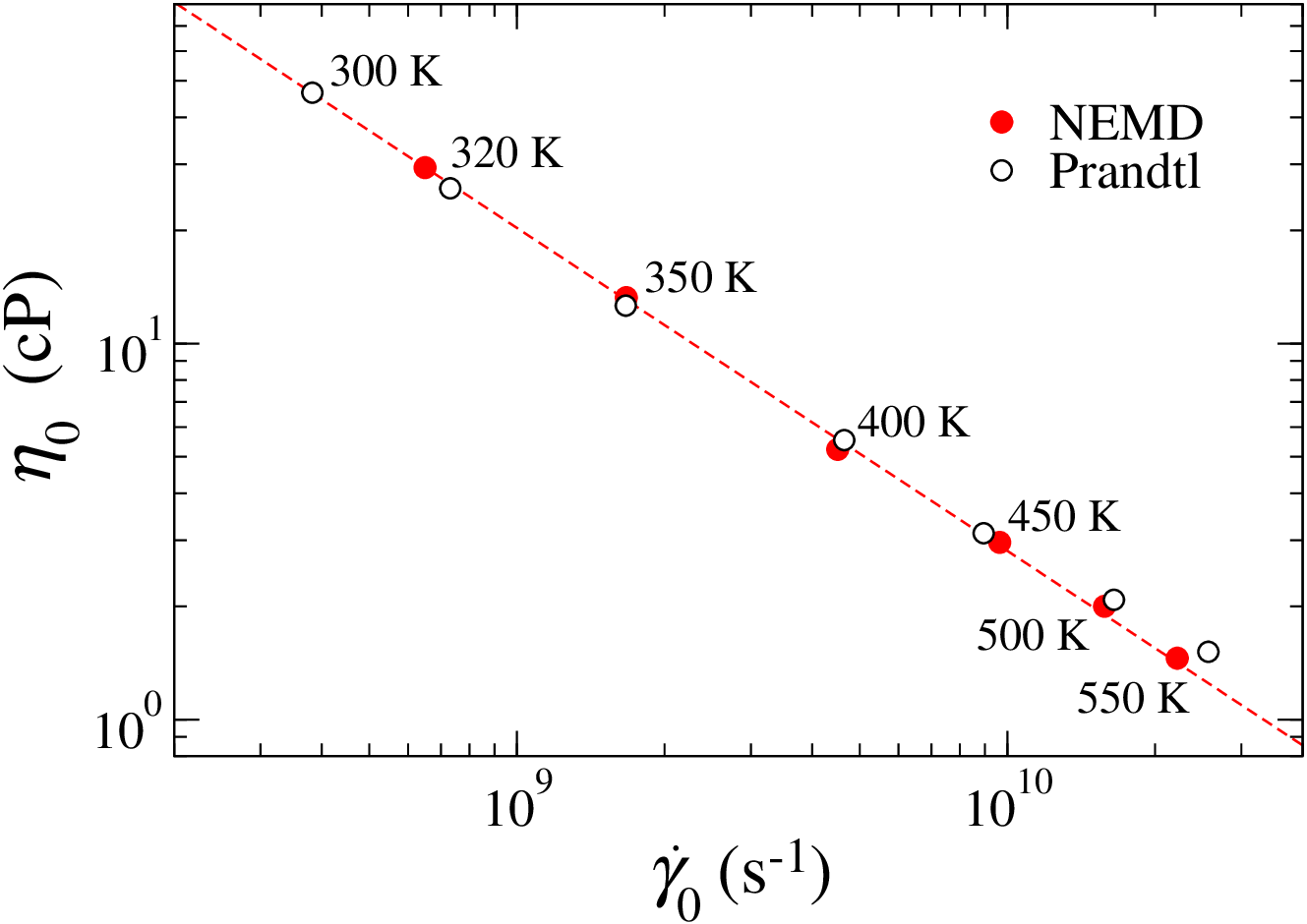}
\caption{\label{fig:prandtl_mapping}
Relation between equilibrium viscosity $\eta_0$ and cross-over shear rate $\dot{\gamma}_0$ as obtained in full NEMD simulations and the Prandtl model.
The dashed line reflects a slightly sublinear, inverse power law, i.e., $\eta_0 \propto 1/\dot{\gamma}^{0.84}$, which is drawn to guide the eye. 
}
\end{figure}

Fig.~\ref{fig:prandtl_mapping} further reveals the large degree of correlation between the results of all-atom simulations and those of the Prandtl model.
Both data sets have the same inverse correlation of $\eta_0$ and $\dot{\gamma}_0$, as would be expected from the equilibrium dynamics of an activated process. 

\begin{figure}[hbtp]
\centering
\includegraphics[width=0.48\textwidth]{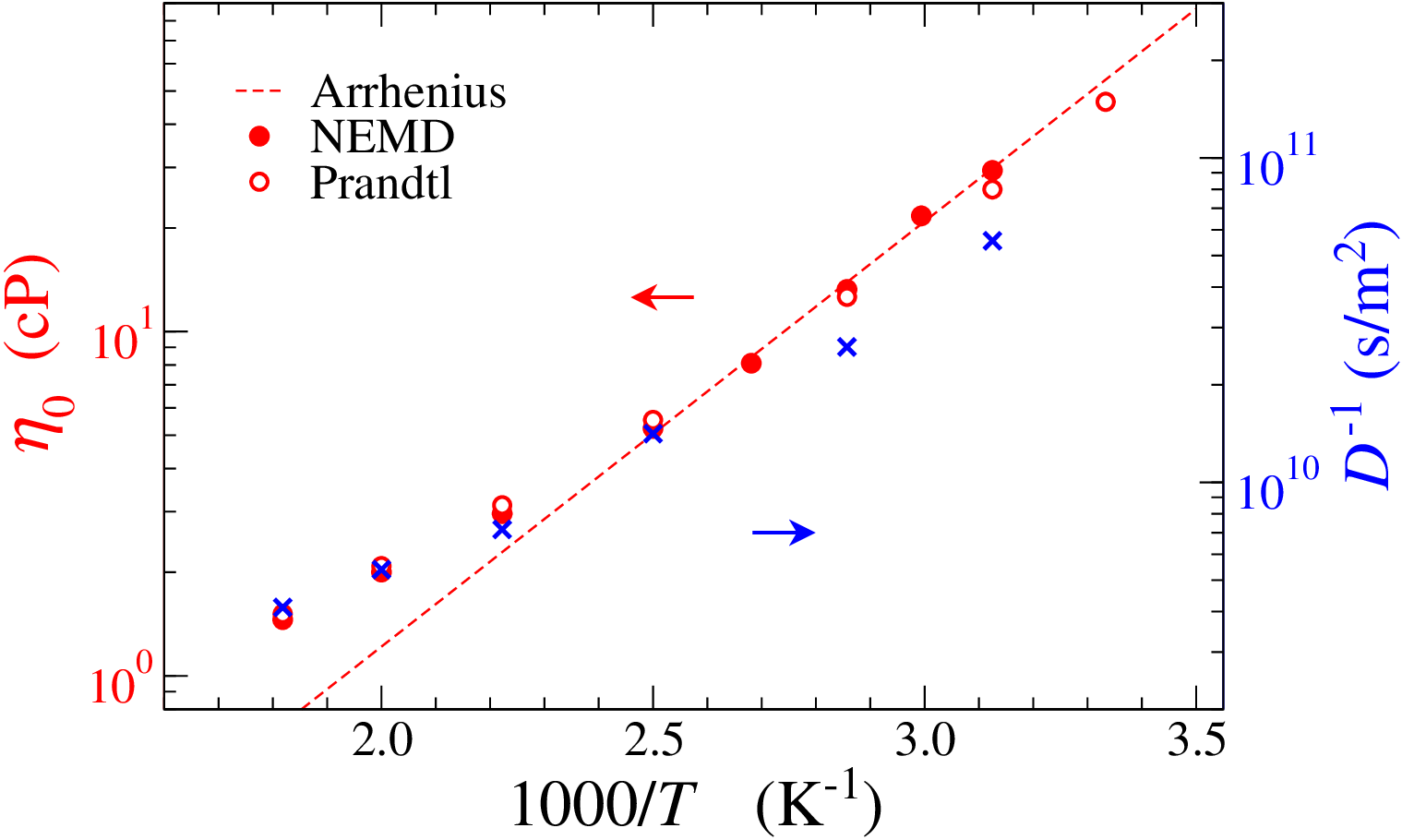}
\caption{ \label{fig:eta0_temp}
Newtonian viscosity $\eta_0$ from all-atom simulations (full, red circles) and Prandtl model (open, black circles) and inverse diffusion coefficient $1/D$ of hexadecane (blue crosses) as a function of inverse temperature, for $n$-hexadecane, at 300~MPa.
Both ordinates, i.e., $\eta_0$ on the left and $1/D$ on the righ, span the same factor of 100. 
}
\end{figure}

The final analysis in this section is concerned with the temperature dependence of hexadecane's viscosity at the relatively large, constant compressive stress of $p_{33}=300$~MPa.
Fig.~\ref{fig:eta0_temp} hints to a cross-over from a non-Arrhenius to an Arrhenius-like dependence in the all-atom MD data.
Unfortunately, some doubt on the robustness of the Arrhenius-type dependence at small temperature remains, as we have not yet managed to compute an equilibrium viscosity below 320~K within our statistical error margin of about 1\% from the full MD simulations.
However, the Prandtl model, which accounts for the cross-over reasonably well, while clearly having a constant energy barrier supports an extension of an Arrhenius dependence to smaller temperature.
The model thereby also reveals that the deviation from Arrhenius at large $T$ might not be caused by temperature-dependent energy barriers but could be owing to the ratio $\Delta E/k_\textrm{B}T$ not being a large number, at which point the exponential factor $\exp(\Delta E/k_\textrm{B}T)$ no longer dominates other temperature dependencies. 

To further illuminate the reason for the FST, we  computed the diffusion coefficient $D$ of carbon atoms from their time-dependent root-mean-square displacement (see also Sect.~\ref{sec:micro}) to explore if the temperature dependence of the diffusion coefficient starts to decouple from that of the viscosity, as it happens at the FST of glass-forming melts.
Fig.~\ref{fig:eta0_temp} provides some evidence for this to happen, which, however is relatively weak, because accurate diffusion coefficients at small temperatures are difficult to determine. 
Thus, more research, which is currently outside the scope of our computational feasibility,  is required to see if (a) the viscosity is really Arrhenius-like at small temperature and (b) if the cross-over can already be rationalized within the Prandtl model or is in need of a more elaborate explanation like the FST.
The difficulty is that statistical error bars are small at relatively small times, at which point the functional form used to fit the time-dependent root-mean-square displacement affect the value for $D$, while stochastic errors are large at large times. 
Thus, values for $1/D$ reported at the lowest temperatures must be taken with a grain of salt and might increase by roughly 20\% with different fitting procedures than those that we assumed. 

%
%
%

%

\subsection{Pressure dependence}

\begin{figure}[hbtp]
\centering
\includegraphics[width=0.4\textwidth]{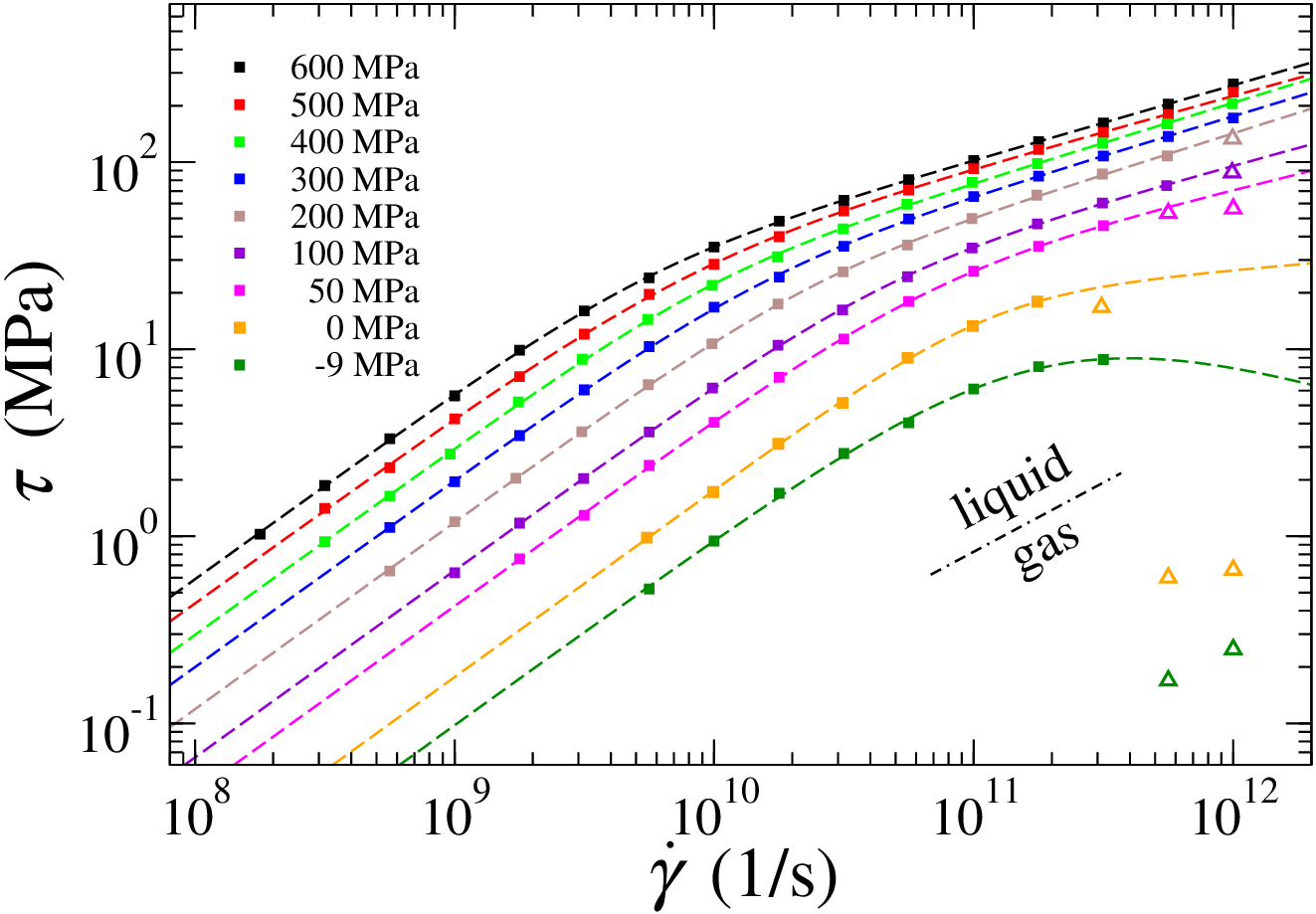}
\caption{\label{fig:eta_gamma_press}
Shear stress $\tau$ of $n$-hexadecane at various pressures (0-600 MPa) and a constant temperature of 500~K. 
Symbols show NEMD data and dashed lines fits to it using the Carreau-Yasuda equation.
At high shear rates, the system can be either in the liquid or in the gas phase.
Open symbols refer to the gas phase as well as data for the liquid phase, if some configurations revealed shear localization. 
}
\end{figure}

Besides the temperature dependence of the viscosity $\eta$ at fixed pressure $p_{33}$, we also analyzed $\eta(p)$ at fixed temperature.
This time, however, we present the results in the form of shear stress $\tau$ as a function of shear rate $\dot{\gamma}$, see the main panel in Fig.~\ref{fig:eta_gamma_press}.
Results are also reported for simulations at a negative normal pressure.
%
They reflect  meta-stable conditions, because a thermally or shear-induced void in the liquid  would unavoidably grow indefinitely under a tensile load once the void surpasses a critical nucleation size. 
However, including tensile stresses into the analysis allows us to explore the pressure-dependence on viscosity at values of $p_{33}$, where this dependence is particularly steep. 

\begin{figure}[ht]
\centering
\includegraphics[width=0.4\textwidth]{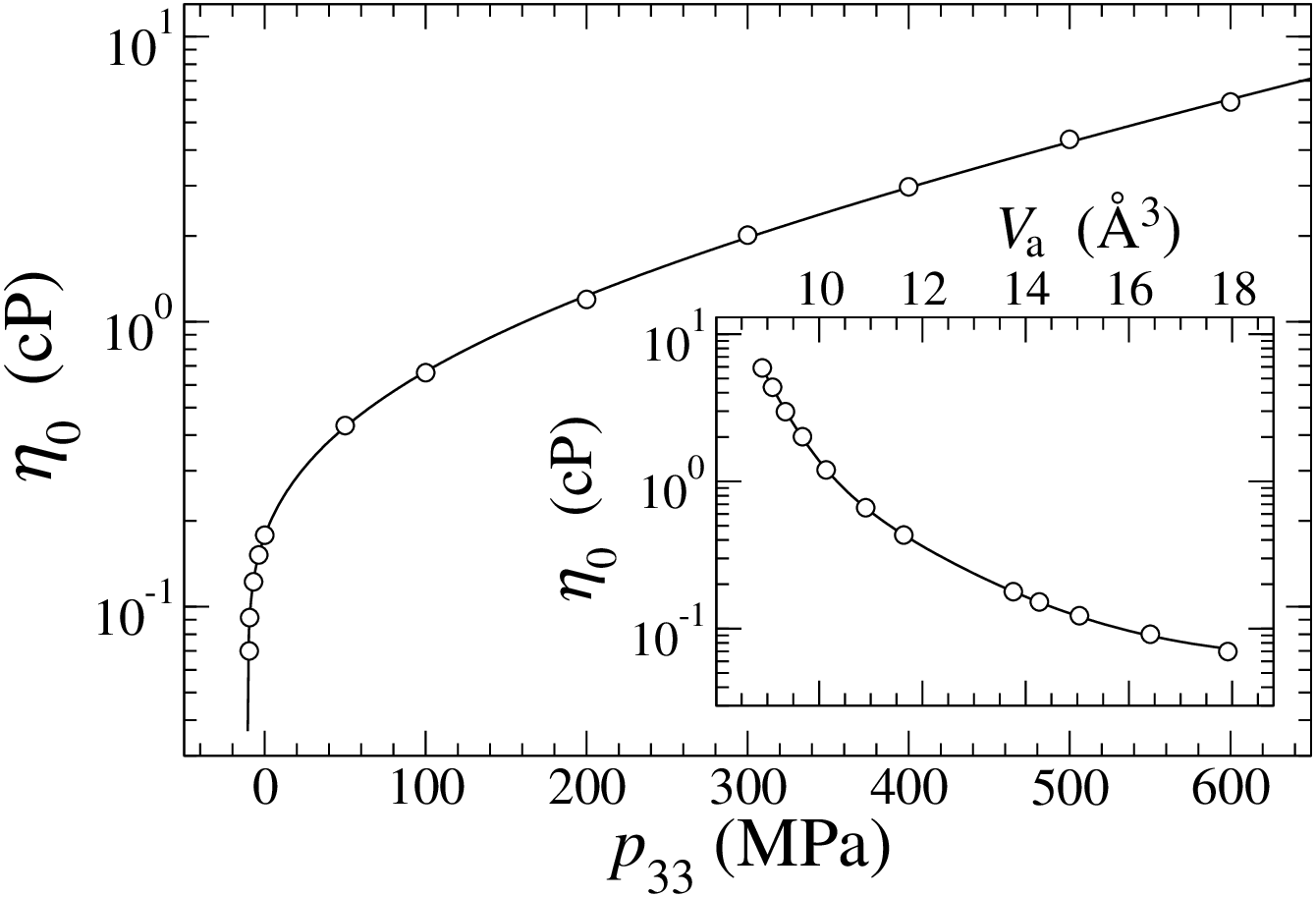}
\caption{Main graph: Newtonian viscosity $\eta_{0}$ as a function of normal pressure $p$, $n$-hexadecane, 500 K.
Symbols show numerical results, while the line is a fit to Eq.~\eqref{eq:eta0_pressure}.
Inset: $\eta_{0}$ as a function of the volume $V$ per atom.
The line is a fourth-order polynomial of $\ln \eta_0$ in $V$.}
\label{fig:eta0_press_vol}
\end{figure}

Analysis of the equilibrium viscosity, see Fig.~\ref{fig:eta0_press_vol}, reveals that $\eta_0$ might disappear as a power law at a negative pressure of $p^* = -10.5$~MPa.
This value is close to the pressure at which the volume of a meta-stable liquid, having been ``stress-quenched'' to a negative pressure, is expected to diverge.
However, at large pressure, $\eta(p)$ crosses over to an exponential dependence of pressure.
A simple functional form reflecting this behavior is
\begin{equation}
\label{eq:eta0_pressure}
    \eta(p) = \eta_1 \left(p/p^*-1\right)^\mu + \eta_2\left(e^{\beta pV^*}-e^{\beta p^*V^*} \right),
\end{equation}
where the parameters $\eta_1$, $\eta_2$, $\mu$, and $V^*$ are phenomenological adjustable parameter and $\beta = 1/k_\textrm{B}T$
$V^*$ can be interpreted as $V^* \approx \partial \Delta E/\partial p$ at intermediate pressures and be called ``free volume'' or Peter, both names probably being similarly meaningful, also because $V^*$ can take a different value in non-equilibrium situations when replacing the hydrostatic pressure $p$ in Eq.~\eqref{eq:eta0_pressure}  with the normal pressure $p_{33}$. 
%
The inset of Fig.~\ref{fig:eta0_press_vol} shows the viscosity as a function of volume. 
No extended domain of a linear $\ln \eta_0$ versus volume can be detected.

A few more words on the decrease of viscosity at $p \gtrsim p^*$ may be in place.
The line represented in Fig.~\ref{fig:eta0_press_vol} is based on an exponent $\mu \approx 2/9$.
However, the uncertainty of the exponent is rather large, as the value changes quite substantially depending on what data are included during the fitting.
(We excluded the point at the lowest pressure.) 
Here, we would only want to state a rough upper bound of 1/2 for the exponent. 
Moreover, we would argue that the decrease in viscosity arises from a much reduced steric repulsion as molecules slide past each other under a tensile hydrostatic stress. 

We do not intend to claim that the functional form of $\eta(p)$ proposed in Eq.~\eqref{eq:eta0_pressure} is superior to other descriptions, in particular generalizations of the McEwen equation~\cite{McEwen1952JIP}, for example by Bair~\cite{Bair2015HTHP}  
or other others discussed by recently by Kondratyuk, Pisarev, and Ewen~\cite{Kondratyuk2020JCP}.
However, our analysis suggests that a quasi-exponential pressure dependence at large compression might be an appropriate description, at least at low shear rates. 

\begin{figure}[ht]
\centering
\includegraphics[width=0.4\textwidth]{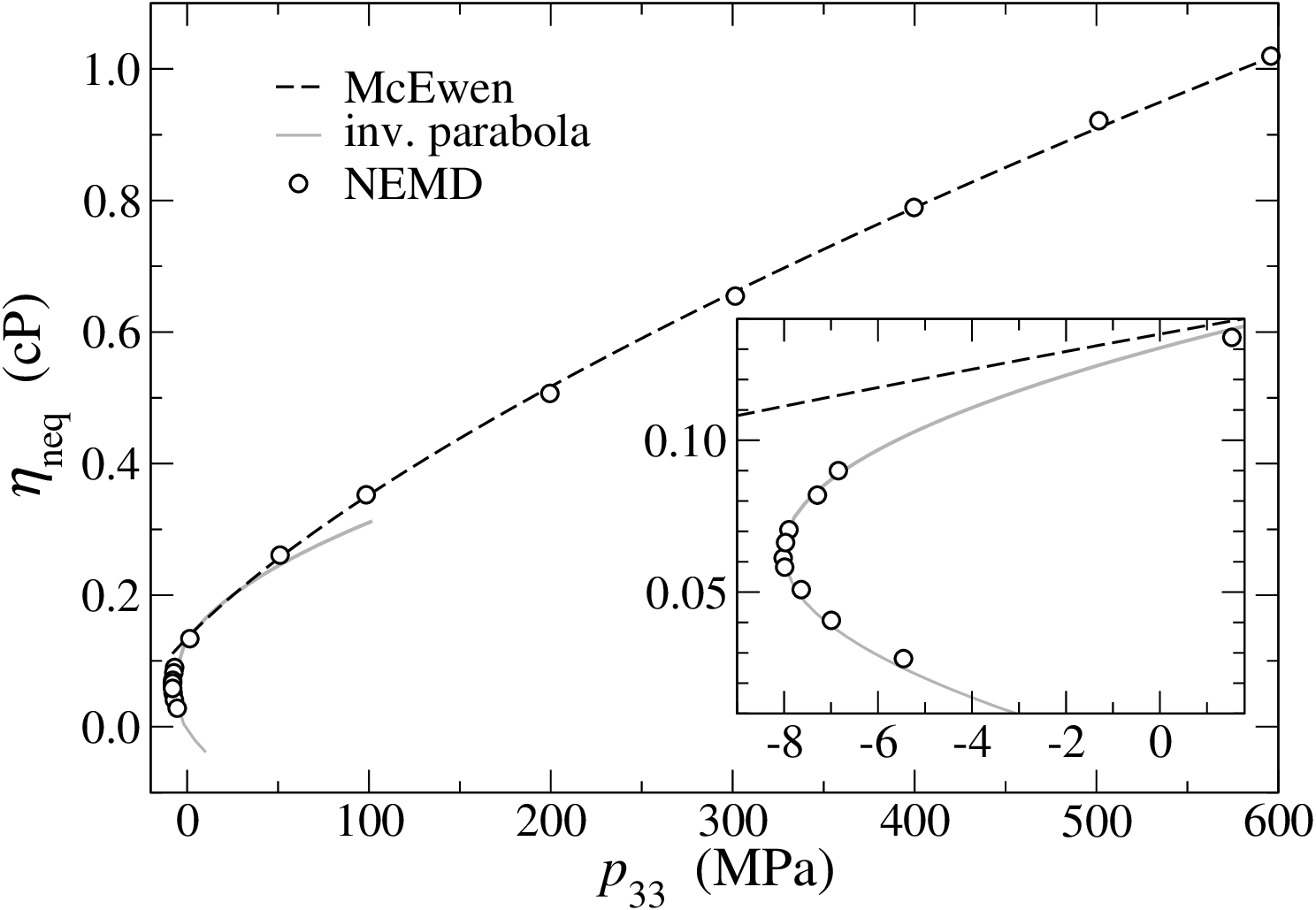}
\caption{Effective viscosity at a high shear rate of $\dot{\nu} = 10^{11}$~Hz at $T = 500$~K as a function of pressure $p_{33}$. The inset highlights result for tensile stresses.}
\label{fig:McEwen}
\end{figure}

In fact, we find the McEwen $\eta(p)$ dependence, which can be cast as the $\eta_2=0$ limiting case of Eq.~\eqref{eq:eta0_pressure}  
to hold extremely well at compressive stress and fixed, large shear rates, as is demonstrated in the main part of Fig.~\ref{fig:McEwen}.
Here $\eta_\textrm{ref}$ is the effective viscosity at a high shear rate and zero normal stress, the exponent $\mu$ is set to $\mu = 1/2$, and $\sigma_0$ a hypothetical tensile stress at which the McEwen formula would predict the viscosity to disappear.
(Using $\mu = 1/2$ substantially worsened the agreement between fit and data.)
However, without further modification, the McEwen equation makes poor predictions once the stress becomes tensile, as can be seen in the inset of Fig.~\ref{fig:McEwen}.

\subsection{Chain-length dependence}

In addition to changing pressure and temperature, we also altered the degree of polymerization, $P$, however, only in a relatively narrow range in which molecules are much too short to entangle. 
From the results reported in Table~\ref{tab:CY_fit_results}, it can be seen that $\eta_0$ and $\dot{\gamma}_0$ scale roughly with $P^2$ and $1/P^2$, respectively.
Changes in the exponents $n$ and $a$ are rather small, which points once more to the existence of an elementary instability. 
This result does not necessarily contradict findings by Sivabeak and Persson (SP)~\cite{Sivebaek2012PRL}, who suggested a more noticeable $n(P)$ dependence, where their symbol $n$ corresponded to our $1-n$.
Further, more substantial difference between their and our study are that
SP explored a relatively small normal pressure of 10~MPa while altering $P$ by a factor of 70, which covers the range from oligomer to strongly entangled polymers. 

An interesting feature born out in our data is that $\eta(\dot{\gamma})$ overlaps at very high shear rates for different degree of polymerization $P$, i.e., not only the exponent $n$ but also the prefactors are similar at high shear rates for different alkanes. 

\begin{figure}[ht]
\centering
\includegraphics[width=0.4\textwidth]{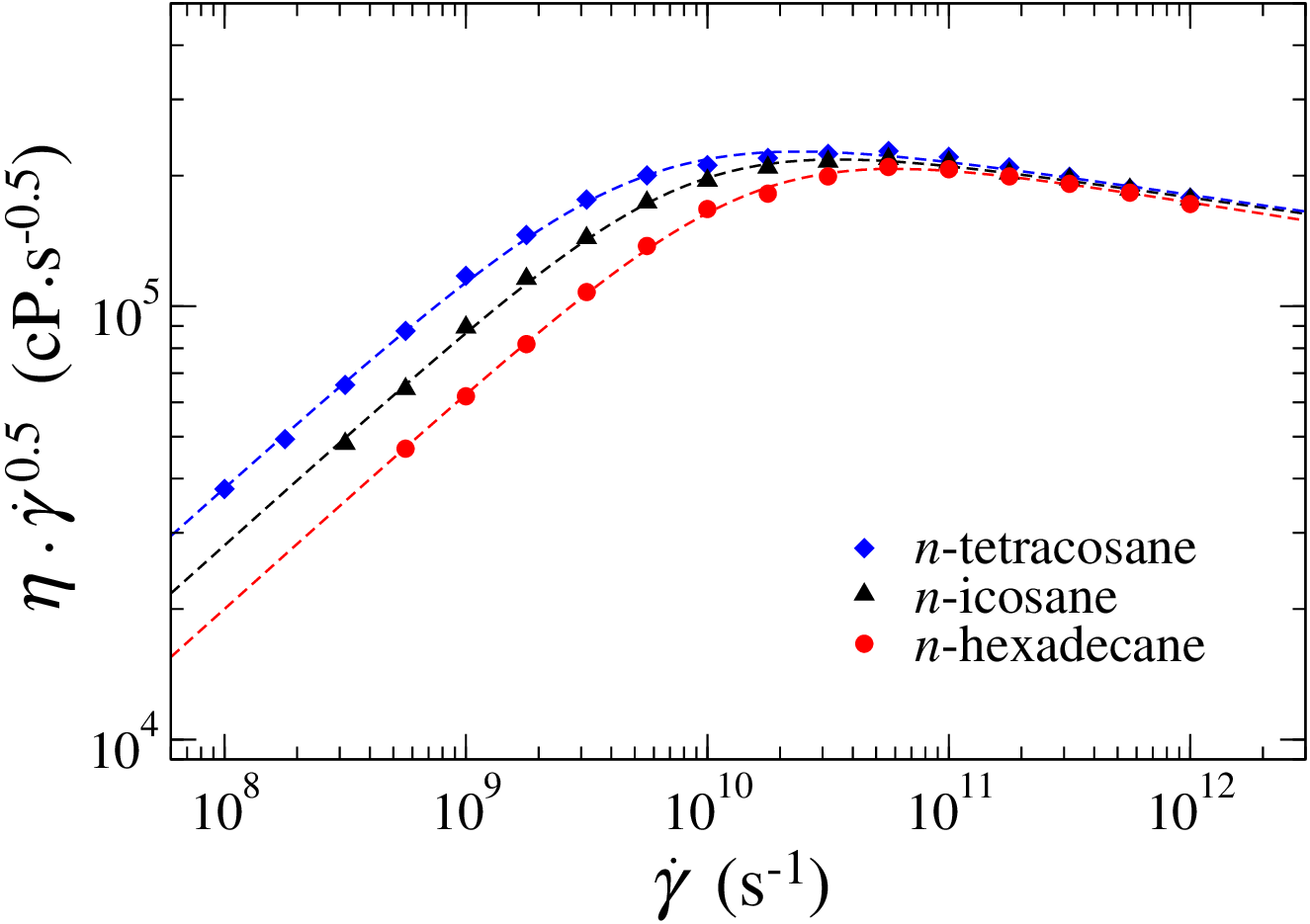}
\caption{Scaled shear stress vs. shear rate for $n$-hexadecane at 500~K and 300~MPa.
}
\label{fig:chem_sigma_gamma}
\end{figure}

\subsection{Microscopic analysis}
\label{sec:micro}

Since the early days of shear-thinning studies, attempts were made to rationalize phenomenological relations in terms of microscopic models and to relate shear thinning to structural changes~\cite{Lacks2001PRL}.
Early theories of polymers in solution~\cite{Lodge1956TFS}, ``derive'' the Carreau equation by considering the rate of formation and breaking of contact points.
Even if the explanation were valid for such systems, it can scarcely matter for dense polymer mixtures, when the resistance to flow is largely affected by steric repulsion.
Yet, non-local configurational changes like polymer alignment have been held responsible for changes in viscosity~\cite{Jadhao2019TL}.
Specifically, polymers oriented along the shear-flow direction were expected to counteract shear in that direction than those with quasi-spherical equilibrium shapes. 

It might have came as a surprise to many when Jadhao and Robbins~\cite{Jadhao2019TL} reported the alignment of polymers to have saturated when the viscosity had merely changed by a factor of two, which, in their master plot was at the very beginning of the  shear-thinning regime.
We can confirm their finding in our work and extend on it by noticing that the molecules tend to coil up again at constant pressure, rather than constant volume, at very high shear rates. 
Thus, Ostwald scaling may occur even if the overall molecular shape develops in a non-monotonic fashion.

\begin{figure}[ht]
\centering
\includegraphics[width=0.4\textwidth]{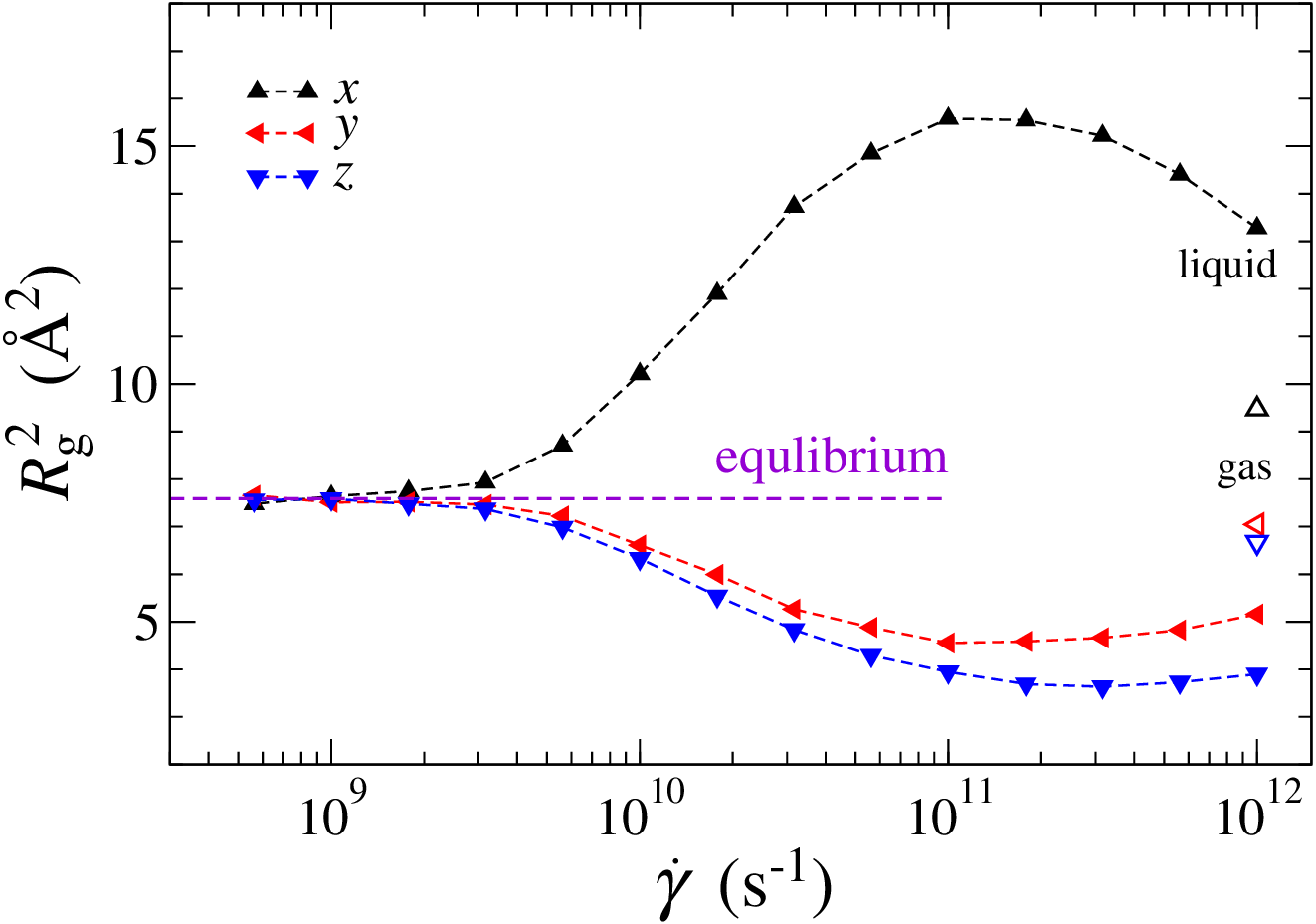}
\caption{Different cartesian components to $R_\textrm{g}^2$ vs shear rate, for $n$-hexadecane at $p = 300$~MPa and $T = 500$~K (full symbols).
Open symbols refer to gas-phase data at $p = 3.4$ MPa.
}
\end{figure}

\begin{figure}[hbtp]
\centering
\includegraphics[width=0.43\textwidth]{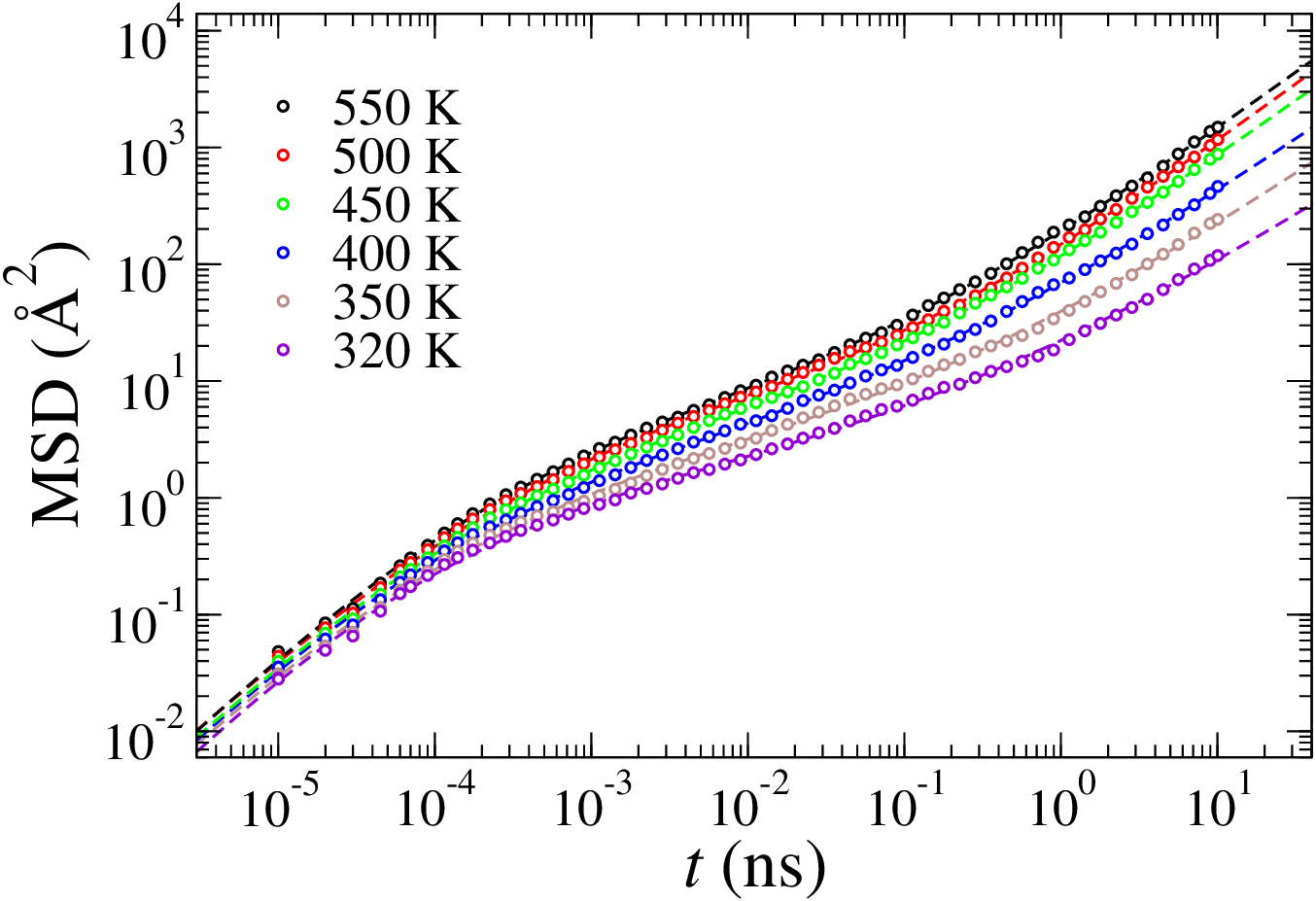}
\caption{Mean-squared displacement (MSD) of $n$-hexadecane from equilibrium MD simulation at 300 MPa. 
Lines are drawn to guide the eye. 
}
\label{fig:eta0_tempXYZ}
\end{figure}

For reason of completeness, we also present data that we produced to determine the diffusion coefficient from equilibrium simulations in Fig.~\ref{fig:eta0_tempXYZ}, which shows the atom averaged mean-square displacement as a function of time. 
Unfortunately, the value for the diffusion coefficient is sensitive to what functional form is assumed.
Assuming $\left\langle \left\{ x(t)-x(0)\right\}^2\right\rangle$ to be a constant plus $D t$ for the large times yields slightly different results than when using the sum of a sub-diffusive and a diffusive function. 
%
%
%
Nonetheless, we note that the time, where sub-diffusive motion crosses over to diffusive motion is of similar order of magnitude as the inverse of the shear rate, at which shear-thinning becomes noticeable. 

\section{Discussion and Conclusions}
\label{sec:conclusions}

This work demonstrates that the shear-thinning of alkanes at high pressure can be mapped quite well onto the Prandtl model.
Fixing its two dimensional parameters at a given temperature and pressure even allows changes in the equilibrium viscosity and the functional form of $\eta(\dot{\gamma})$ to be reproduced over a reasonably broad temperature range without readjusting parameters.
To a large degree, the parametrization is even transferable between different alkanes. 
Moreover, the obtained activation barrier of $\Delta E\approx 0.08$~eV that we obtain at a pressure of $p=300$~MPa is greater but of similar order of magnitude as experimental result for the relaxational dynamics in polyethylen melts at zero pressure, which are close to 0.05~eV at ambient pressure~\cite{Qiu1999M}.
Despite this success, we do not claim our parametrization to be optimal but rather minimal in the sense that the smallest number of adjustable parameters needed to describe shear-thinning was used.
In principle, it could be attempted to also target the linear-response frequency dependence $\eta(\omega)$.

Our simulation data on all-atom alkanes, in particular $n$-hexadecane also indicates that the additional parameter(s) used in a Carreau-Yasuda fit are justified, at least at elevated pressure.
We find a fit to have a median error of roughly 12\%, which reduces to about 6\% when using an additional parameter in the Cross model.
This error is cut once more into one half when using Carreau instaed of Cross, and by almost another factor of two with Carreau-Yasuda, at which point the systematic error is on par with our stochastic error. 







\section*{Acknowledgments}
We thank Jörg Baschnagel and Daniele Dini for useful discussions. 

\section*{Declarations}
\subsection*{Funding} 
This research was supported by the German Research Foundation (DFG) under grant number GA 3059/2-1. 

\subsection*{Conflict of interest} No.

\subsection*{Authors' contributions} 
MHM designed and supervised the project, wrote the manuscript.
HG carried out all the simulations, analyzed the data and prepared the figures.
MHM and HG reviewed and modified the manuscript.

\bibliographystyle{plain}
\bibliography{reference}

\begin{thebibliography}{10}

\bibitem{Bair2017PNAS}
S.~Bair.
\newblock Purported fragile-to-arrhenius crossover in squalane.
\newblock {\em Proc. Natl. Acad. Sci. U.S.A.}, 114(42), October 2017.

\bibitem{Bair2003TT}
S.~Bair and P.~Kottke.
\newblock Pressure-viscosity relationships for elastohydrodynamics.
\newblock {\em Tribol. Trans.}, 46(3):289--295, January 2003.

\bibitem{Bair2015TL}
S.~Bair, P.~Vergne, P.~Kumar, G.~Poll, I.~Krupka, M.~Hartl, W.~Habchi, and
  R.~Larsson.
\newblock Comment on {\textquotedblleft}history, origins and prediction of
  elastohydrodynamic friction{\textquotedblright} by spikes and jie.
\newblock {\em Tribol. Lett.}, 58(1), March 2015.

\bibitem{Bair2015HTHP}
Scott Bair.
\newblock Choosing pressure-viscosity relations.
\newblock {\em High Temperatures-High Pressures}, 44:415–428, 2015.

\bibitem{Bair2002PRL}
Scott Bair, Clare McCabe, and Peter~T. Cummings.
\newblock Comparison of nonequilibrium molecular dynamics with experimental
  measurements in the nonlinear shear-thinning regime.
\newblock {\em Phys. Rev. Lett.}, 88:058302, Jan 2002.

\bibitem{Barnes1997JNNFM}
Howard~A Barnes.
\newblock Thixotropy{\textemdash}a review.
\newblock {\em Journal of Non-Newtonian Fluid Mechanics}, 70(1-2):1--33, May
  1997.

\bibitem{Carreau1972TSR}
Pierre~J. Carreau.
\newblock Rheological equations from molecular network theories.
\newblock {\em Transactions of the Society of Rheology}, 16(1):99--127, March
  1972.

\bibitem{Crook1963MPS}
A.~W. Crook and Thomas~Edward Allibone.
\newblock The lubrication of rollers iv. measurements of friction and effective
  viscosity.
\newblock {\em Philosophical Transactions of the Royal Society of London.
  Series A, Mathematical and Physical Sciences}, 255(1056):281--312, 1963.

\bibitem{Daivis1994JCP}
Peter~J. Daivis and Denis~J. Evans.
\newblock {Comparison of constant pressure and constant volume nonequilibrium
  simulations of sheared model decane}.
\newblock {\em The Journal of Chemical Physics}, 100(1):541--547, 01 1994.

\bibitem{deAlmeidaRibeiro2020PRR}
Ingrid de~Almeida~Ribeiro and Maurice de~Koning.
\newblock Non-newtonian flow effects in supercooled water.
\newblock {\em Phys. Rev. Res.}, 2(2), April 2020.

\bibitem{Ewen2019PCCP}
James~P. Ewen, Hongyu Gao, Martin~H. Müser, and Daniele Dini.
\newblock Shear heating{,} flow{,} and friction of confined molecular fluids at
  high pressure.
\newblock {\em Phys. Chem. Chem. Phys.}, 21:5813--5823, 2019.

\bibitem{Ewen2021TL}
James~P. Ewen, Hugh~A. Spikes, and Daniele Dini.
\newblock Contributions of molecular dynamics simulations to elastohydrodynamic
  lubrication.
\newblock {\em Tribology Letters}, 69:24, 2021.

\bibitem{Eyring2004JCP}
Henry Eyring.
\newblock {Viscosity, Plasticity, and Diffusion as Examples of Absolute
  Reaction Rates}.
\newblock {\em The Journal of Chemical Physics}, 4(4):283--291, 12 1936.

\bibitem{Fisher1985PRB}
Daniel~S. Fisher.
\newblock Sliding charge-density waves as a dynamic critical phenomenon.
\newblock {\em Physical Review B}, 31(3):1396--1427, February 1985.

\bibitem{Jadhao2017PNAS-R}
V.~Jadhao and M.~O. Robbins.
\newblock Reply to bair: Crossover to arrhenius behavior at high viscosities in
  squalane.
\newblock {\em Proc. Natl. Acad. Sci. U.S.A.}, 114(42), 2017.

\bibitem{Jadhao2017PNAS}
Vikram Jadhao and Mark~O. Robbins.
\newblock Probing large viscosities in glass-formers with nonequilibrium
  simulations.
\newblock {\em Proc. Natl. Acad. Sci. U.S.A.}, 114(30):7952--7957, 2017.

\bibitem{Jadhao2019TL}
Vikram Jadhao and Mark~O. Robbins.
\newblock Rheological properties of liquids under conditions of
  elastohydrodynamic lubrication.
\newblock {\em Tribology Letters}, 67:66, 2019.

\bibitem{Kondratyuk2020JCP}
Nikolay~D. Kondratyuk, Vasily~V. Pisarev, and James~P. Ewen.
\newblock {Probing the high-pressure viscosity of hydrocarbon mixtures using
  molecular dynamics simulations}.
\newblock {\em The Journal of Chemical Physics}, 153(15):154502, 10 2020.

\bibitem{Lacks2001PRL}
Daniel~J. Lacks.
\newblock Energy landscapes and the non-newtonian viscosity of liquids and
  glasses.
\newblock {\em Phys. Rev. Lett.}, 87:225502, Nov 2001.

\bibitem{Lees1972JPC}
A~W Lees and S~F Edwards.
\newblock The computer study of transport processes under extreme conditions.
\newblock {\em Journal of Physics C: Solid State Physics}, 5(15):1921, aug
  1972.

\bibitem{Lemarchand2015JCP}
Claire~A. Lemarchand, Nicholas~P. Bailey, Billy~D. Todd, Peter~J. Daivis, and
  Jesper~S. Hansen.
\newblock {Non-Newtonian behavior and molecular structure of Cooee bitumen
  under shear flow: A non-equilibrium molecular dynamics study}.
\newblock {\em The Journal of Chemical Physics}, 142(24):244501, 06 2015.

\bibitem{Li2000MSE}
R~Li.
\newblock Time-temperature superposition method for glass transition
  temperature of plastic materials.
\newblock {\em Mater. Sci. Eng.}, A278:36--45, 2000.

\bibitem{Lodge1956TFS}
A.~S. Lodge.
\newblock A network theory of flow birefringence and stress in concentrated
  polymer solutions.
\newblock {\em Trans. Faraday Soc.}, 52:120--130, 1956.

\bibitem{Lucas2021JNCS}
Pierre Lucas.
\newblock Fragile-to-strong transitions in glass forming liquids.
\newblock {\em J. Non-Cryst. Solids}, 557:119367, April 2021.

\bibitem{McEwen1952JIP}
E.~McEwen.
\newblock The effect of variation of viscosity with pressure on the
  load-carrying capacity of the oil film between gear-teeth.
\newblock {\em J. Inst. Pet.}, 38(1):646--672, 1952.

\bibitem{Muser2011PRB}
Martin~H. M\"{u}ser.
\newblock Velocity dependence of kinetic friction in the prandtl-tomlinson
  model.
\newblock {\em Physical Review B}, 84(12), September 2011.

\bibitem{Muser2020L}
Martin~H. M\"{u}ser.
\newblock Shear thinning in the \protect{Prandtl} model and its relation to
  generalized \protect{Newtonian} fluids.
\newblock {\em Lubricants}, 8(4):38, March 2020.

\bibitem{Ostwald1925KZ}
Wolfgang Ostwald.
\newblock Ueber die geschwindigkeitsfunktion der viskosit\"{a}t disperser
  systeme. i.
\newblock {\em Kolloid-Zeitschrift}, 36(2):99--117, February 1925.

\bibitem{Plimpton1995JCP}
Steve Plimpton.
\newblock Fast parallel algorithms for short-range molecular dynamics.
\newblock {\em Journal of Computational Physics}, 117(1):1--19, 1995.

\bibitem{Popov2012ZAMM}
V.L. Popov and J.A.T. Gray.
\newblock Prandtl-tomlinson model: History and applications in friction,
  plasticity, and nanotechnologies.
\newblock {\em {ZAMM} - Journal of Applied Mathematics and Mechanics /
  Zeitschrift f\"{u}r Angewandte Mathematik und Mechanik}, 92(9):683--708, July
  2012.

\bibitem{Prandtl1928ZAMM}
L.~Prandtl.
\newblock Ein gedankenmodell zur kinetischen theorie der festen körper.
\newblock {\em J. Appl. Math. Mech.}, 8(2):85--106, 1928.

\bibitem{Price2001JCC}
Melissa L.~P. Price, Dennis Ostrovsky, and William~L. Jorgensen.
\newblock Gas-phase and liquid-state properties of esters, nitriles, and nitro
  compounds with the opls-aa force field.
\newblock {\em Journal of Computational Chemistry}, 22(13):1340--1352, 2001.

\bibitem{Qiu1999M}
XiaoHua Qiu and M.~D. Ediger.
\newblock Local and global dynamics of unentangled polyethylene melts by
  $^{13}$c nmr.
\newblock {\em Macromolecules}, 33(2):490--498, December 1999.

\bibitem{Roland2006JCP}
C.~M. Roland, S.~Bair, and R.~Casalini.
\newblock Thermodynamic scaling of the viscosity of van der waals, h-bonded,
  and ionic liquids.
\newblock {\em J. Chem. Phys.}, 125(12):124508, September 2006.

\bibitem{Ruina1983JGR}
Andy Ruina.
\newblock Slip instability and state variable friction laws.
\newblock {\em Journal of Geophysical Research: Solid Earth},
  88(B12):10359--10370, December 1983.

\bibitem{Schalek1923KZ}
E.~Schalek and A.~Szegvari.
\newblock Die langsame koagulation konzentrierter eisenoxydsole zu reversiblen
  gallerten.
\newblock {\em Kolloid-Zeitschrift}, 33(6):326--334, December 1923.

\bibitem{Siu2012JCTC}
Shirley W.~I. Siu, Kristyna Pluhackova, and Rainer~A. Böckmann.
\newblock Optimization of the opls-aa force field for long hydrocarbons.
\newblock {\em Journal of Chemical Theory and Computation}, 8(4):1459--1470,
  2012.
\newblock PMID: 26596756.

\bibitem{Sivebaek2012PRL}
I.~M. Sivebaek, V.~N. Samoilov, and B.~N.~J. Persson.
\newblock Effective viscosity of confined hydrocarbons.
\newblock {\em Physical Review Letters}, 108(3):036102, January 2012.

\bibitem{Smith1996PRB}
Elizabeth~D. Smith, Mark~O. Robbins, and Marek Cieplak.
\newblock Friction on adsorbed monolayers.
\newblock {\em Physical Review B}, 54(11):8252--8260, September 1996.

\bibitem{Spikes2014TL}
H.~Spikes and Z.~Jie.
\newblock History, origins and prediction of elastohydrodynamic friction.
\newblock {\em Tribol. Lett.}, 56(1):1--25, August 2014.

\bibitem{Spikes2015TL-R}
H.~Spikes and W.~Tysoe.
\newblock On the commonality between theoretical models for fluid and solid
  friction, wear and tribochemistry.
\newblock {\em Tribol. Lett.}, 59(1), June 2015.

\bibitem{Spikes2015TL}
H.~Spikes and J.~Zhang.
\newblock Reply to the comment by scott bair, philippe vergne, punit kumar,
  gerhard poll, ivan krupka, martin hartl, wassim habchi, roland larson on
  {\textquotedblleft}history, origins and prediction of elastohydrodynamic
  friction{\textquotedblright} by spikes and jie in tribology letters.
\newblock {\em Tribol. Lett.}, 58(1), March 2015.

\bibitem{Spikes2017TL}
H.~A. Spikes.
\newblock Comment on: Rheology of an ionic liquid with variable carreau
  exponent: A full picture by molecular simulation with experimental
  contribution, by nicolas voeltzel, philippe vergne, nicolas fillot, nathalie
  bouscharain, laurent joly, tribology letters (2016) 64:25.
\newblock {\em Tribol. Lett.}, 65(2), April 2017.

\bibitem{Todd2017}
Billy~D. Todd and Peter~J. Daivis.
\newblock {\em Nonequilibrium Molecular Dynamics: Theory, Algorithms and
  Applications}.
\newblock Cambridge University Press, 2017.

\bibitem{Voeltzel2016TL}
N.~Voeltzel, P.~Vergne, N.~Fillot, N.~Bouscharain, and L.~Joly.
\newblock Rheology of an ionic liquid with variable carreau exponent: A full
  picture by molecular simulation with experimental contribution.
\newblock {\em Tribol. Lett.}, 64(2), October 2016.

\bibitem{Voeltzel2017TL}
Nicolas Voeltzel, Philippe Vergne, Nicolas Fillot, Nathalie Bouscharain, and
  Laurent Joly.
\newblock Reply to the {\textquotedblleft}comment on `rheology of an ionic
  liquid with variable carreau exponent: A full picture by molecular simulation
  with experimental contribution, ' by n. voeltzel, p. vergne, n. fillot, n.
  bouscharain, l. joly, tribology letters (2016) 64:25{\textquotedblright} by
  h. a. spikes.
\newblock {\em Tribol. Lett.}, 65(2), April 2017.

\end{thebibliography}

\end{document}